\documentclass[prb,twocolumn,superscriptaddress,amsmath,amssymb,floatfix,preprintnumbers]{revtex4-1}
\usepackage{epsfig}
\usepackage{amsmath}
\usepackage{graphicx}
\usepackage{placeins}
\usepackage{float}
\usepackage{dcolumn}
\usepackage{color}
\usepackage{lipsum}
\usepackage{epsfig}  
\usepackage{epstopdf}
\usepackage{physics} 
\usepackage{bm}
\usepackage[all]{xy}
\usepackage{braket}
\DeclareGraphicsExtensions{.png .jpg .pdf}

\usepackage{dsfont}
\newcommand*{\hham}{\mathcal{H}}
\newcommand{\angstrom}{\text{\normalfont\AA}}
\usepackage{mathtools} 
\usepackage{hyperref}
\hypersetup{colorlinks = true,linkcolor = blue,anchorcolor = blue,citecolor = blue,filecolor = blue,urlcolor = blue}

\begin{document}

\title{Electronic properties of multilayered Lieb, transition, and Kagome lattices}

\author{T. F. O. Lara }\email{temersonlara@fisica.ufc.br}
\affiliation{Departamento de F\'isica, Universidade Federal do Cear\'a, Campus do Pici, 60455-900 Fortaleza, Cear\'a, Brazil}

\author{E. B. Barros}
\affiliation{Departamento de F\'isica, Universidade Federal do Cear\'a, Campus do Pici, 60455-900 Fortaleza, Cear\'a, Brazil}

\author{W. P. Lima}
\affiliation{Departamento de F\'isica, Universidade Federal do Cear\'a, Campus do Pici, 60455-900 Fortaleza, Cear\'a, Brazil}

\author{J. P. G. Nascimento} \affiliation{Departamento de F\'isica, Universidade Federal do Cear\'a, Campus do Pici, 60455-900 Fortaleza, Cear\'a, Brazil}

\author{J. Milton Pereira Jr.}
\affiliation{Departamento de F\'isica, Universidade Federal do Cear\'a, Campus do Pici, 60455-900 Fortaleza, Cear\'a, Brazil}

\author{T. A. S. Pereira}
\affiliation{Instituto de F\'isica, Universidade Federal de Mato Grosso, 78060-900, Cuiab\'a, Mato Grosso, Brazil}
\affiliation{National Institute of Science and Technology on Materials Informatics, Campinas, Brazil}

\author{D. R. da Costa}
\email{diego\_rabelo@fisica.ufc.br}
\affiliation{Departamento de F\'isica, Universidade Federal do Cear\'a, Campus do Pici, 60455-900 Fortaleza, Cear\'a, Brazil}

\begin{abstract}
Based on the interconvertibility feature shared between monolayer Lieb and Kagome lattices, which allows mapping transition lattice's stages between these two limits ($\pi/2 \leq\theta \leq 2\pi/3$), in this work we extend the recently proposed one-control ($\theta$) parameter tight-binding model for the case of a multilayer Lieb-Kagome system, by considering the two most-common stacks: AA and AB (Bernal). We systematically study the band transformations between the two lattices by adjusting the interlayer hopping and distance, with or without considering the influence of the nearest interlayer neighbors, for different numbers of stacked layers, and under the application of an external perpendicular electric field. The energetic changes are understood from the perspective of the layer dependence of the pseudospin components and the total probability density distributions. The present framework provides an appropriate and straightforward theoretical approach to continuously investigate the evolution of electronic properties in the multilayer Lieb-Kagome system under various external effects.
\end{abstract}

\maketitle

\section{Introduction}

The search for new materials with unique physical properties represents one of the main areas of focus in current research in condensed matter physics and materials engineering. \cite{lei2022graphene, ago2024science, shanmugam2022review} Among them, flat-band lamellar materials are a class of the two-dimensional (2D) family that stands out \cite{wang2022intrinsic, PhysRevB.101.045131, yin2022topological, PhysRevB.107.184441, PhysRevMaterials.5.084203, huang2023recent} as a consequence of their unusual phenomena, such as zero group velocities and suppressed transport, making them sensitive to external interactions and favoring correlated phases, such as ferromagnetism and superconductivity~\cite{doi:10.1080/23746149.2018.1473052, PhysRevLett.109.067201, Liu_2014}. Notable examples include Lieb~\cite{SADEGHI2023115172, chen2017spin, PhysRevB.82.085310} and Kagome~\cite{PhysRevB.80.193304, PhysRevB.80.113102, mojarro2023strain} lattices, which combine flat bands and Dirac cones in their electronic structure.

\begin{figure*}
{\includegraphics[width=0.9\linewidth]{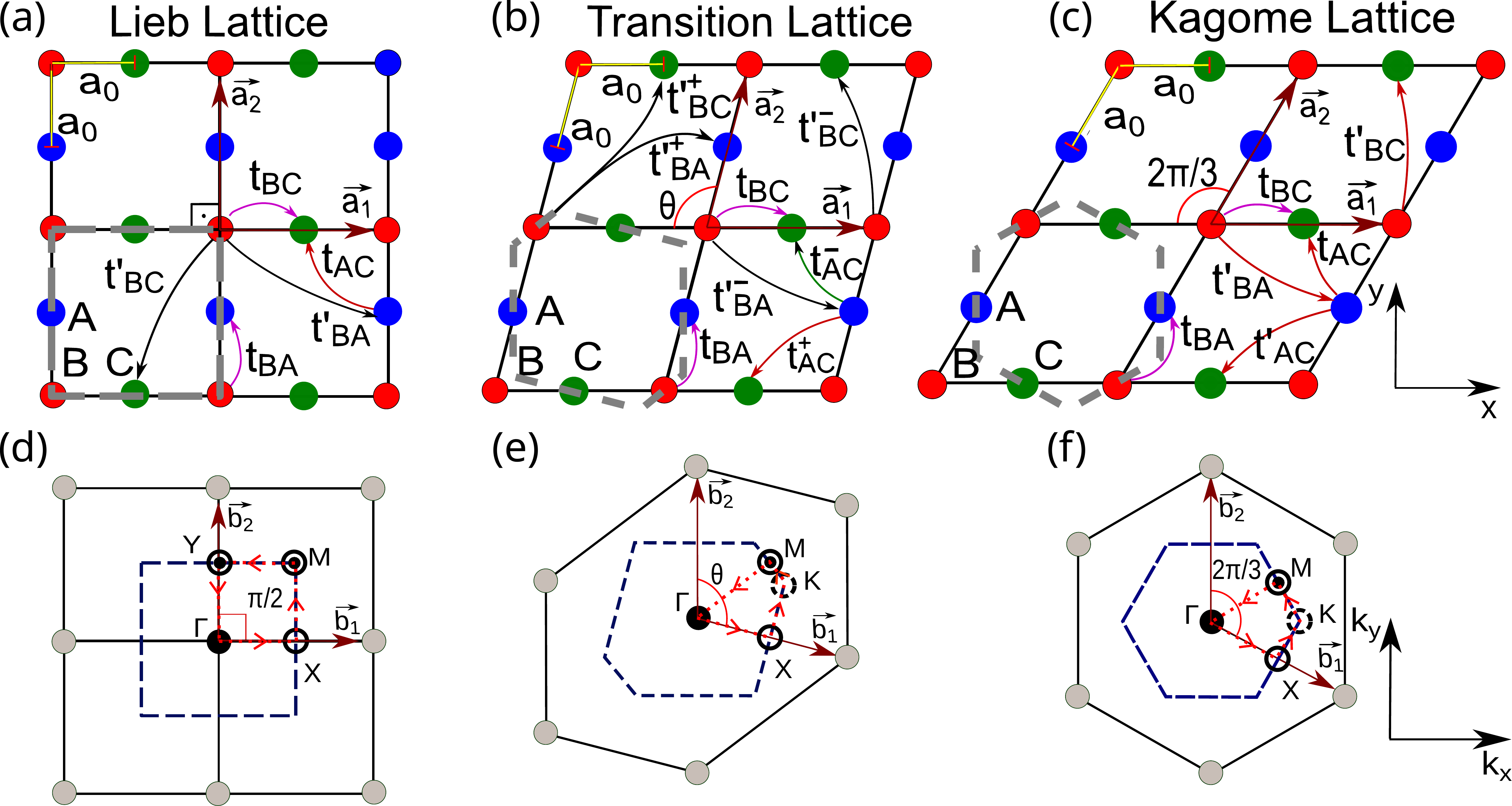}}
\caption{\textcolor{blue}{(Color online)} Real (upper panels) and reciprocal (lower panels) generic lattices: (a) Lieb lattice - $\hat{D}_{4h}$ ($\theta=\pi/2$), (b) transition lattice - $\hat{D}_{2h}$ ($\pi/2<\theta<2\pi/3$) and (c) Kagome lattice - $\hat{D}_{6h}$ ($\theta=2\pi/3$). $\vec{a}_1$ and $\vec{a}_2$ (brown arrows) are the primitive vectors with unit cells denoted by the gray dashed lines containing three nonequivalent sites A (blue circle), B (red circle), and C (green circle). The distance and hopping parameters between the nearest neighboring sites are $a_0$, $t_{BA}$, and $t_{BC}$. $t_{AC}^{\pm}, \;t_{BA}'^{\pm}$ and $t_{BC}'^{\pm}$ correspond to the first, second, and third neighbors, respectively. First Brillouin zone of the (d) Lieb, (e) transition, and (f) Kagome lattices are denoted by blue dashed lines. The reciprocal vectors are $\vec{b}_1$ and $\vec{b}_2$ (brown arrows) and the high-symmetry points are also shown: $\vec{\Gamma}$ (filled circle), $\vec{X}$ (empty solid circle), $\vec{M}$ (circle with dot inside), $\vec{K}$ (empty dashed circle) and $\vec{Y}$ (dashed circle with dot inside).}
\label{real_lattices}
\end{figure*}

The Lieb lattice is a 2D periodic system with three sites (labeled here as A, B, and C -- see Fig.~\ref{real_lattices}(a)) per unit cell. Excitations in such systems are known to present spectra with Dirac cones and flat bands~\cite{SADEGHI2023115172, chen2017spin, PhysRevB.82.085310}. In the case of electrons, the existence of flat bands grants singular properties, influencing the charge carriers' response to external magnetic fields~\cite{aoki1996hofstadter,goldman2011topological,nictua2013spectral}, and optics~\cite{han2022optical} and transport~\cite{jakubsky2023lieb, PhysRevB.96.024304} phenomena. Due to these characteristics, Lieb lattice has been explored experimentally in photonic lattices~\cite{vicencio2015observation,mukherjee2015observation}, ultracold atoms~\cite{goldman2011topological, PhysRevB.81.041410}, and organometallic materials~\cite{jiang2020topological,jiang2021exotic,cui2020realization}. The Kagome lattice, like the Lieb lattice, consists of three sites (labeled here as A, B, and C -- see Fig.~\ref{real_lattices}(c)) per unit cell, where A and C occupy the center of the edges, and B the lattice structure vertices. Its band structure features a flat band and two dispersive bands forming a Dirac cone, similar to graphene~\cite{PhysRevB.80.113102, PhysRevLett.69.1608,kimura2002magnetic}. The Kagome lattice exhibits properties remarkably distinct from those observed in the Lieb lattice \cite{kimura2002magnetic, PhysRevB.80.193304,xiao2003landau, PhysRevB.98.245145,zhu2017shot,Pereira_2010}, and its structure is most commonly found in organometallic~\cite{PhysRevB.96.115426,crasto2019layertronic,kumar2021manifestation,wang2023p} and magnetic systems~\cite{ye2018massive, liu2018giant, nakatsuji2015large}.

In 2019, Jiang \textit{et al.}\cite{tony2019} demonstrated that, through diagonal strains, it is possible to convert the Lieb lattice into the Kagome lattice, observing that the flat band of the Lieb lattice continuously migrates to the lower part of the Kagome spectrum. Recent works have explored this interconvertibility both theoretically and experimentally. \cite{PhysRevB.101.045131, PhysRevA.107.023509, wang2023p, cui2020realization, jiang2021exotic, PhysRevB.108.125433, uchoa2025electronic, wellissonTPT} For example, Lim \textit{et al.}\cite{PhysRevB.101.045131} used a tight-binding model to investigate the evolution of Dirac points during the transition, modifying the hopping between sites A and C, while Lang \textit{et al.}\cite{PhysRevA.107.023509} studied tilted Dirac cones in this transitional lattice, demonstrating their observation in photonic lattices by means of simulations and experimental implementations.

Layered material systems are formed by reducing the dimensionality of bulk (three-dimensional, 3D) systems to 2D architectures. \cite{wang2024stacking, chaves2020bandgap, pham2024layer} The electronic properties of these multilayer systems are critically influenced by their ordering, composition, and relative chemical bonds' angle, \textit{i.e.}, the relative orientation of their constituent layers. This adjustment enables different material stackings,\cite{guo2021stacking,bonaccorso2012production} such as AA, AB, and ABC, influencing their physical properties and response to external fields. \cite{PhysRevB.107.045429, doi:10.1143/JPSJ.76.024701, doi:10.1143/JPSJ.76.024701, C5CP05013H, Min2012} For instance, for AA-stacked graphene, its band structure exhibits shifted Dirac cones, whose separation increases under an electric field \cite{ould2017electronic, ROZHKOV20161}, whereas in Bernal (AB) stacked configuration, its dispersion relation is quadratic, and a perpendicular electric field opens an energy gap. \cite{PhysRevB.84.195453, ROZHKOV20161, McCann_2013, Castro_2010} The thickness-dependent physics is not a particularity of multilayered graphene; two relevant examples are the transition metal dichalcogenides (TMDs) that are semiconductors whose bandgap transition varies between monolayers and thicker structures\cite{wang2012electronics,splendiani2010emerging,terrones2014bilayers,he2014stacking}, and phosphorene that has an adjustable energy gap depending on the number of layers and electric field, expanding its applications in optoelectronics and nanophotonics. \cite{PhysRevB.96.155427, PhysRevB.89.201408, PhysRevB.99.235424, PhysRevB.103.165428}

In the context of multilayer systems, the electronic properties of the Lieb lattice change significantly depending on the type of stacking and the number of layers. Banerjee \textit{et al.}\cite{banerjee2021higher} studied the topological properties of the Lieb lattice in AA, AB, and ABC stacking, observing that in the AB and ABC cases, the energy spectra exhibit characteristics of nonsymmorphic symmetry. Regarding multilayer Kagome lattice systems, several studies in layer degree of freedom engineering have been conducted in organometallic systems~\cite{PhysRevB.96.115426} and twisted systems~\cite{sinha2021twisting, wu2021bilayer, PhysRevB.100.155421}.

Thus, motivated by the above scenario of systems formed by multilayers of 2D materials and the exploration of interconvertibility between Lieb and Kagome lattices, we systematically investigate in the present work the impact of two stacking types (AA and AB) on the energy spectrum and the effects of a perpendicular electric field on multilayer Lieb-Kagome systems. For that, we organize the paper as follows. The theoretical framework for dealing with the monolayer Lieb-Kagome lattice is presented in Sec.~\ref{sec.I}. To be more specific, in Sec.~\ref{sec.crystal}, we introduce the crystallographic aspects (such as the real and reciprocal lattice vectors and the high-symmetry points of the corresponding reciprocal lattices, all of them with a $\theta$ dependence) of the monolayer Lieb-Kagome lattice, which enables the general study of both Lieb and Kagome lattices as a unified structure. The general tight-binding model for the Lieb-Kagome lattice in terms of the interconversion parameter $\theta$ is derived in Sec.~\ref{sec.TB}. Discussions on the band structure evolution between Lieb and Kagome lattices as well as the main electronic features of monolayer Lieb (Sec.~\ref{sec.mono.lieb}), transition (Sec.~\ref{sec.mono.transition}), and Kagome (Sec.~\ref{sec.mono.kagome}) lattices are pointed out. The multilayer system model for the AA and AB stacking is presented in Sec.~\ref{sec.II}, first deriving the bilayer case (Sec.~\ref{sec.bilayer}) and then extending the theoretical formulation to the multilayer system (Sec.~\ref{sec.multilayer}). Sections~\ref{sec.III} and \ref{sec.IV} are devoted to the results and discussions in the presence and absence of a perpendicular electric field for AA-stacked and AB-stacked  Lieb-Kagome lattices, respectively. Finally, our main findings are summarized in Sec.~\ref{sec.conclusions}.

\section{Monolayer Lieb-Kagome lattice}\label{sec.I}

\subsection{Crystallographic aspects}\label{sec.crystal}

The generic Lieb-Kagome lattice is defined as a 2D crystalline structure with a unit cell composed of three sites at the basis (labeled by A, B, and C), which may present symmetry $\hat{D}_{4h}$, $\hat{D}_{2h}$ and $\hat{D}_{6h}$ according to the angle choice $\angle{ABC}$, ranging from $\pi/2\leq\theta\leq2\pi/3$. As represented in the Figs.~\ref{real_lattices}(a) and \ref{real_lattices}(c), the limit cases of $\theta=\pi/2$ and $\theta=2\pi/3$ correspond to the Lieb ($\hat{D}_{4h}$) and Kagome ($\hat{D}_{6h}$) lattices, respectively. The intermediate 2D crystals obtained with $\pi/2<\theta<2\pi/3$ are called transition lattices ($\hat{D}_{2h}$), as shown in Fig.~\ref{real_lattices}(b), since they represent the evolution stages between Lieb and Kagome lattices in the interconversion process, as well discussed in the Refs.~[\onlinecite{tony2019, Cui2019, osti_1527138, PhysRevB.101.045131, wellissonTPT}].

The generic primitive lattice vectors are
\begin{equation}\label{prim_geral}
\vec{a}_1=a\hat{\upsilon}_1, \quad\mbox{and} \quad \vec{a}_2=a\hat{\upsilon}_2,
\end{equation} 
with $\hat{\upsilon}_1=(1, 0)$ and $\hat{\upsilon}_2=\left(-\cos\theta,\sin\theta\right)$. In order to investigate a more generic case, we assume, without loss of generality, that $a=1$ \angstrom. The reciprocal lattice, in turn, is generated by the following primitive vectors
\begin{equation}\label{eq b generico}
\vec{b}_1=b_1\hat{\nu}_1, \quad \mbox{and}  \quad \vec{b}_2=b_2\hat{\nu}_2,
\end{equation}
with $|\vec{b}_1|=|\vec{b}_2|={2\pi}/{(a\sin\theta)}\equiv b$, $\hat{\nu}_1=\left(\sin\theta,\cos\theta\right)$ and $\hat{\nu}_2=\left(0,1\right)$. The high-symmetry points in the first Brillouin zone are given by
\begin{eqnarray}\label{pontos_reciproca_lieb_kagome}
&&\vec{\Gamma}=(0,0),\qquad \vec{X}=\frac{b}{2}\hat{\nu}_1, \qquad\vec{M}=\frac{b}{2}\left(\hat{\nu}_1+\hat{\nu}_2\right),
\nonumber\\
&& 
\vec{K}=\frac{1}{2}\left(b-l\cos\theta\right)\hat{\nu_1}+\frac{l}{2}\hat{\nu}_2,\qquad \vec{Y}=\frac{b}{2}\hat{\nu}_2,
\end{eqnarray}
with $l={b}/(1-\cos\theta)$. In Figs.~\ref{real_lattices}(d)--\ref{real_lattices}(f), the reciprocal space of the Lieb, transition, and Kagome lattices, respectively, with the first Brillouin zone and the high-symmetry points are presented. Throughout this work, we will use the generic lattice to simultaneously study Lieb-Kagome lattices stacked in a multilayer system by means of a generic $\theta$-dependent tight-binding model, as described in the next section.  

\subsection{Tight-binding model}\label{sec.TB}

The tight-binding Hamiltonian of a single-layer Lieb-Kagome lattice can be given by 
\begin{equation}\label{eqh1}
    \hat{\hham}_{sl} = \sum_{i}\varepsilon_{i}c_{i}^{\dagger}c_{i} +\sum_{i\neq j}t_{ij}c_{i}^{\dagger}c_{j},
\end{equation}
where $i$ and $j$ run over the lattice sites, $t_{ij}$ is hopping parameter between the $i$ and $j$ sites, and $c_{i}^{\dagger}\;(c_{j})$ is the creation (annihilation) operator of electrons at site $i\;(j)$, and $\varepsilon_{i}$ is the on-site energy at sublattice $i=\{A, B, C\}$.

Since the unit cell of the generic lattice has three distinct sites ($A$, $B$, and $C$), as depicted in Fig.~\ref{real_lattices}(b), the Hamiltonian of Eq.~\eqref{eqh1} can be rewritten as 
\begin{align}\label{eqh2}
\hat{\hham}_{sl}& = \sum_{j}\left(  E_A{{\hat{a}_j}}^{\dagger}\hat{a}_j +E_B{\hat{b}_{j}^{\dagger}}\hat{b}_{j}+E_C{\hat{c}_{j}^{\dagger}}\hat{c}_{j}\right)\nonumber\\
& + \sum_{ij}\left[{t_{BA}}\left({\hat{b}_{j}^{\dagger}}\hat{a}_{i} + {\hat{a}_{i}}^{\dagger}\hat{b}_{j}\right) +{t_{BC}}\left({\hat{b}_{j}^{\dagger}}\hat{c}_{i} \hspace{-0.1cm}+ \hspace{-0.1cm}{\hat{c}_{i}}^{\dagger}\hat{b}_{j}\right)\right]\nonumber\\
&+\sum_{ij}\left[{t_{AC}}\left({\hat{a}_{j}^{\dagger}}\hat{c}_{i}+{\hat{c}_{i}}^{\dagger}\hat{a}_{j}\right)\right],
\end{align}
where $\hat{a}_{j}$ (${\hat{a}_{j}}^{\dagger}$), $\hat{b}_{j}$ (${\hat{b}_{j}}^{\dagger}$) and $\hat{c}_{j}$ (${\hat{c}_{j}}^{\dagger}$) are annihilation (creation) operators corresponding to the $j$th-site of sublattices $A$, $B$, and $C$, respectively. Due to the in-plane translational symmetry, one can Fourier transform the field operators, such as \cite{lima2022tight}
\begin{equation}\label{fourie}
    \hat{\alpha}_{j}^{\dagger} = \frac{1}{\sqrt{N}}\sum_{k}e^{-i\vec{k}\cdot\vec{r}_j}\hat{\alpha}_{k}^{\dagger},\;\;\textup{and}\;\;\hat{\alpha}_{j} = \frac{1}{\sqrt{N}}\sum_{k}e^{i\vec{k}\cdot\vec{r}_j}\hat{\alpha}_{k},
\end{equation}
replacing them, with $\hat{\alpha}_{j}$ denoting $\hat{\alpha}_{j}\equiv \{\hat{a}_{j}, \hat{b}_{j}, \hat{c}_{j} \}$, into Eq.~\eqref{eqh2}, resulting in
\begin{align}\label{eqh3}
\hat{\hham}_{sl}& = \sum_{k}\hspace{-0.05cm}\left(  E_A{{\hat{a}_k}}^{\dagger}\hat{a}_k +E_B{\hat{b}_{k}^{\dagger}}\hat{b}_{k}\hspace{-0.075cm}+E_C{\hat{c}_{k}^{\dagger}}\hat{c}_{k}\right)\nonumber\\
& + \sum_{k}\left [ t_{BA}\sum_{j}e^{-\vec{k}\cdot\vec{\delta}_{ij}}\hat{b}_{k}^{\dagger} \hat{a}_{k} + t_{BA}\sum_{i}e^{-\vec{k}\cdot\vec{\delta}_{ij}}\hat{a}_{k}^{\dagger} \hat{b}_{k} \right ]\nonumber\\
& + \sum_{k}\left [ t_{BC}\sum_{j}e^{-\vec{k}\cdot\vec{\delta}_{ij}}\hat{b}_{k}^{\dagger} \hat{c}_{k} + t_{BC}\sum_{i}e^{-\vec{k}\cdot\vec{\delta}_{ij}}\hat{c}_{k}^{\dagger} \hat{b}_{k} \right ]\nonumber\\
& + \sum_{k}\left [ t_{AC}\sum_{j}e^{-\vec{k}\cdot\vec{\delta}_{ij}}\hat{a}_{k}^{\dagger} \hat{c}_{k} + t_{AC}\sum_{i}e^{-\vec{k}\cdot\vec{\delta}_{ij}}\hat{c}_{k}^{\dagger} \hat{a}_{k} \right ],
\end{align}
where $\vec{\delta}_{ij}$ is the relative position between the sites located at $i$ and $j$. In our model, we consider the hopping parameter amplitudes to be finite between the first $\left ( B \leftrightarrow A,  B \leftrightarrow C \right )$, second $\left ( A \leftrightarrow C\right )$, and third $\left ( B \leftrightarrow A,  B \leftrightarrow C \right )$ nearest-neighbor sites. For the first nearest neighbor, the $\vec{\delta}_{ij}$ are located on $\pm\vec{a}_2/2$ and $\pm\vec{a}_1/2$, the second nearest neighbor on $\pm\left(\vec{a}_1-\vec{a}_2\right)/2$ and $\pm\left(\vec{a}_1+\vec{a}_2\right)/2$, and the third nearest neighbor on $\pm\left(\vec{a}_1+\vec{a}_2/2\right)$, $\pm\left(\vec{a}_1-\vec{a}_2/2\right)$, $\pm\left(\vec{a}_1/2+\vec{a}_2\right)$, and $\pm\left(\vec{a}_1/2-\vec{a}_2\right)$.

Then, Lieb-Kagome band structure is obtained by solving the Schrödinger equation, $\hat{\hham}_{sl}\ket{\Psi} = E\ket{\Psi}$, \textit{i.e.} applying the Hamiltonian operator \eqref{eqh3} to the eigenstate of the system, that in second quantization can be written as
\begin{equation}\label{eqh4}
    \ket{\Psi} = \sum_{k}\left(\phi_{a} {\hat{a}_{k}}^{\dagger} + \phi_{b} {\hat{b}_{k}}^{\dagger} + \phi_{c} {\hat{c}_{k}}^{\dagger} \right)\ket{0},
\end{equation}
where $\phi_{a}$, $\phi_{b}$, and $\phi_{c}$ are the probability amplitudes for finding electrons on the sites $A$, $B$, and $C$, respectively. Using the anti-commutation relations of the creation and annihilation operators, such as $\left \{ \alpha_{\vec{k}},\beta_{\vec{k}} \right \} = 0$ and $\left \{ \alpha_{\vec{k}},\beta_{{\vec{k}}'}^{\dagger} \right \} = \delta_{\vec{k}{\vec{k}}'}\delta_{\alpha \beta}$ where $\{\alpha,\beta\} = \{a,b,c\}$, the eigenvalue problem for a state $\vec{k}$ can be written in the following form
\begin{equation}\label{Sch}
\hham_{k}\psi_{k} = E \psi_{k},
\end{equation} 
with the Hamiltonian for the state $\vec{k}$ given by
\begin{equation}\label{eqh5}
\hat{\hham}_{\vec{k}} = \left( \begin{array}{rccr}
{\hham}_{AA} & {\hham}_{BA} & {\hham}_{AC}\\ 
& {\hham}_{BB} & {\hham}_{BC}\\
& & {\hham}_{CC}\\
\end{array} \right), 
\end{equation} 
whose off-diagonal matrix elements are
\begin{subequations}\label{eqh6}
	\begin{align}
	\hham_{BA}(\vec{k})&=2t_{BA}\cos\left(a_0\vec{k}\cdot\hat{\upsilon}_2\right)\nonumber\\ 
    &+ 2{t}'^{-}_{BA}\cos\left(2a_0\vec{k}\cdot(\hat{\upsilon}_1-\hat{\upsilon}_2/2)\right)\nonumber\\&+ 2{t}'^{+}_{BA}\cos\left(2a_0\vec{k}\cdot(\hat{\upsilon}_1+\hat{\upsilon}_2/2)\right),\\
	\hham_{BC}(\vec{k})&=2t_{BC}\cos\left(a_0\vec{k}\cdot\hat{\upsilon}_1\right)\nonumber\\ &+ 2{t}'^{-}_{BC}\cos\left(2a_0\vec{k}\cdot(\hat{\upsilon}_1/2-\hat{\upsilon}_2)\right)\nonumber\\&+ 2{t}'^{+}_{BC}\cos\left(2a_0\vec{k}\cdot(\hat{\upsilon}_1/2+\hat{\upsilon}_2)\right),\\
	\hham_{AC}(\vec{k})&=2{t_{AC}^{-}}\cos\left[a_0\vec{k}\cdot(\hat{\upsilon}_1-\hat{\upsilon}_2)\right]\nonumber\\
&+2{t_{AC}^{+}}\cos\left[a_0\vec{k}\cdot(\hat{\upsilon}_1+\hat{\upsilon}_2)\right],
	\end{align}
\end{subequations}
and the main diagonal matrix elements ${\hham}_{\alpha\alpha}(\vec{k}) = E_{\alpha}$, with $\alpha\equiv \{A, B, C\}$, are related to the on-site energies of sites A, B, and C. The omitted lower triangle of the matrix in Eq.~\eqref{eqh5} should be filled in according to a hermitian matrix. Furthermore, the eigenstates of the Hamiltonian \eqref{eqh5} are three-component spinors, such as $\psi_{\vec{k}} = \left(\phi_{a}\;\;\;\;  \phi_{b}\;\;\;\;  \phi_{c}\right)^{T}$. Since $\hat{\upsilon}_{1,2} \equiv \hat{\upsilon}_{1,2}(\theta)$ carries out the explicit $\theta$ dependence, it is evident from matrix elements given by Eq.~\eqref{eqh6} that the tight-binding Hamiltonian for the generic lattice [Eq.~\eqref{eqh5}] is also a function of the interconvertibility angle $\theta$, which allows mapping transition lattice's stages between these two limits, \textit{i.e.} for $\pi/2 < \theta < 2\pi/3$. For $\theta = \pi/2$ [$\theta = 2\pi/3$], the Hamiltonian \eqref{eqh5} describes the single-layer Lieb [Kagome] lattice, in which the hopping parameters are ${t}_{BA} = {t}_{BC} \equiv  t$ [${t}_{BA} = {t}_{BC} = {t}_{AC}^{-}  \equiv  t$], ${t}_{AC}^{-} = {t}_{AC}^{+} \equiv t'$ [${t}_{AC}^{+} = {t}'^{-}_{BA} = {t}'^{-}_{BC}\equiv t'$], and ${t}'^{-}_{BC} = {t}'^{+}_{BC} = {t}'^{-}_{BA}={t}'^{+}_{BA} \equiv {t}''$ [${t}'^{+}_{BC}={t}'^{+}_{BA} \equiv {t}''$], which correspond to the first ($t$), second ($t'$), and third ($t''$) nearest neighbors connections, respectively. By neglecting terms associated with third neighbors, the matrix elements in Eq.~\eqref{eqh6} become precisely the ones for the Hamiltonian reported in Ref.~[\onlinecite{tony2019}] used to study the interconvertibility between the Lieb and Kagome lattices for an unstrained generic lattice.

To account for the distance dependence on the hopping parameters, we adopt the hopping normalization described by Lima \textit{et al.} \cite{PhysRevB.108.125433, wellissonTPT} and given by
\begin{equation}\label{hopping1}
    t_{ij} = t\frac{a_{0}}{a_{ij}} e^{-n\left (a_{ij}/a_{0} - 1  \right )},
\end{equation}
in which a dimensionless multiplicative term $a_0 /a_{ij}$ is considered, where $a_{ij}$ the interatomic distance between sites $i$ and $j$, being $a_{0}$ and $t$ the nearest-neighbor distance and hopping parameter, respectively. The parameter $n$ determines the hopping energy's decay ratio as a distance function. Here, we adopt that Eq.~\eqref{hopping1} governs both the intralayer and interlayer hopping parameters, regardless of the stack type. The value of $n$ is set equal to $8$ to provide quasi-flat bands in the energy spectrum and a smooth transition of the band structure between the Lieb and Kagome lattices as proposed by Jiang \textit{et al.} \cite{tony2019}. Lima \textit{et al.} \cite{PhysRevB.108.125433, wellissonTPT} performed a detailed discussion about the choice of the $n$-parameter and its effect on the monolayer Lieb-Kagome band structure (see Sec.~SV in the Supplemental Material of Ref.~[\onlinecite{PhysRevB.108.125433}]). An additional discussion of the hopping energies on the distance dependence in the matter of covering distances that take first, second, or third neighbors is presented in Sec.~\textcolor{blue}{SI} of the Supplemental Materials \cite{SI}.

\subsection{Lieb lattice}\label{sec.mono.lieb}

\begin{figure}[h!]
    \centering
    \includegraphics[scale=0.21]{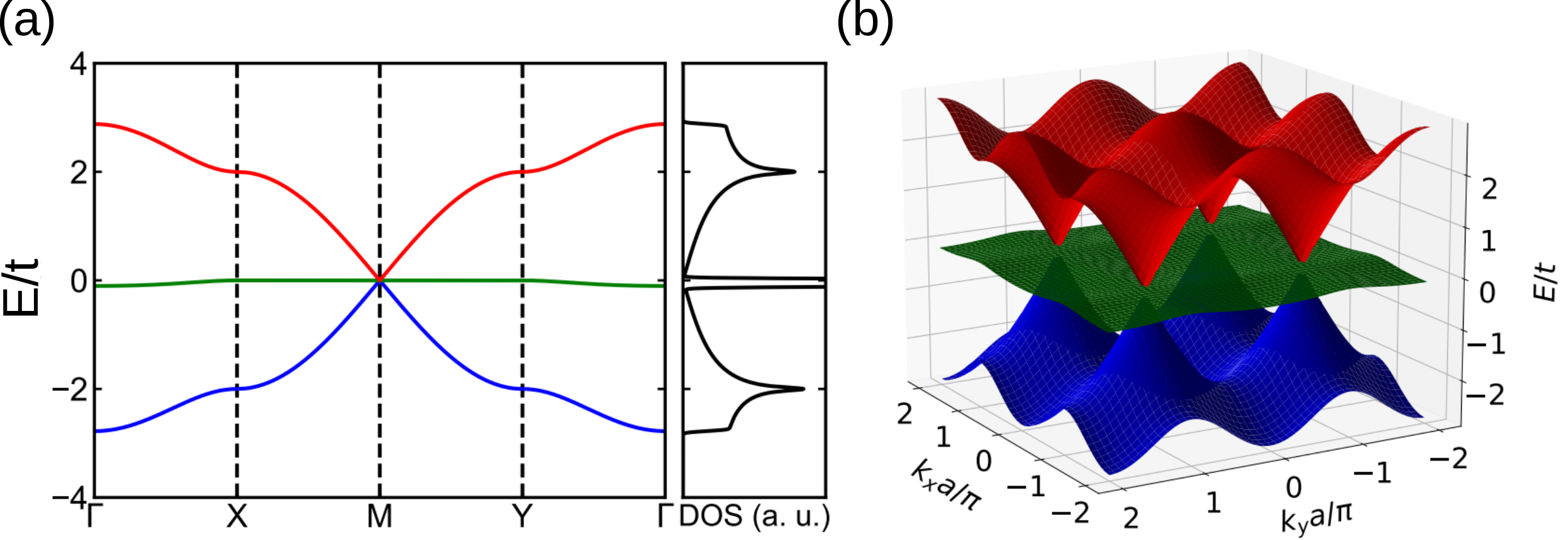}
    \caption{Band structure for the Lieb lattice ($\theta = \pi/2$), obtained by tight-binding calculations, presented in (a) the 2D format along the path connecting the high-symmetry points $\vec{\Gamma}-\vec{X}-\vec{M}-\vec{Y}-\vec{\Gamma}$, and in (b) the 3D surface plot format. The density of states is also shown in (a).}
    \label{Espectro_lieb}
\end{figure}

By diagonalizing the Hamiltonian of Eq.~\eqref{eqh5} with $\theta = \pi/2$, one obtains the energy spectrum for the Lieb lattice. Note in Fig.~\ref{Espectro_lieb} that the energy spectrum for the Lieb lattice comprises two dispersive bands, forming the Dirac cone at the point $\vec{M}$ in the first Brillouin zone, and a third flat band that crosses the Dirac cone \cite{chen2017spin, PhysRevB.82.085310}. The density of states is also presented in Fig.~\ref{Espectro_lieb}(a), in which a pronounced peak characteristic of the flat band and two smaller peaks close to $E/t = \pm 2$ due to van-Hove singularities are observed. Figure~\ref{Espectro_lieb}(b) shows the 3D energy spectrum throughout the reciprocal space, where two dispersive bands forming four Dirac cones are observed, in addition to a flat band that crosses these cones forming triply degenerate points.

\subsection{Transition lattice}\label{sec.mono.transition}

 \begin{figure}[h!]
    \centering
    \includegraphics[scale=0.21]{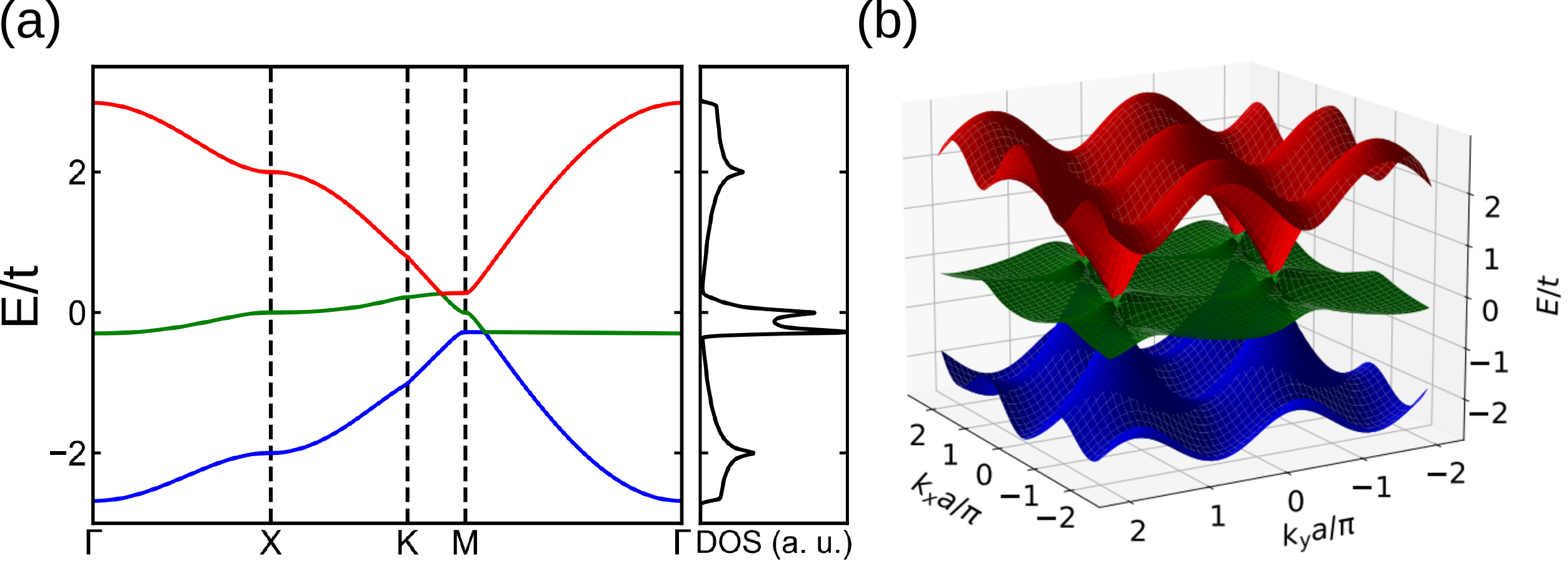}
    \caption{The same as in Fig.~\ref{Espectro_lieb}, but now for the transition lattice taking $\theta = 7\pi/12$. The path connecting the high-symmetry points is now $\vec{\Gamma}-\vec{X}-\vec{K}-\vec{M}-\vec{\Gamma}$.}
    \label{fEspectro_Transicao_mon}
\end{figure}

Figure~\ref{fEspectro_Transicao_mon} shows the band structure for the transition lattice with \(\theta = 7\pi/12\): in Fig.~\ref{fEspectro_Transicao_mon}(a), the 2D format along the high-symmetry points and in Fig.~\ref{fEspectro_Transicao_mon}(b) in the 3D format. They were obtained by diagonalizing the Hamiltonian \eqref{eqh5} with $\theta = 7\pi/12$. In the interconversion process, the deformation of the flat band is observed throughout the reciprocal space, without any \emph{gap} opening between the bands, remaining connected with the upper and lower bands. The triply degenerate Lieb point, $\vec{M}$, splits into two inclined Dirac cones, classified as type I and type II \cite{tony2019, PhysRevA.107.023509}, as shown in Fig.~\ref{fEspectro_Transicao_mon}(a). As $\theta$ increases, one of the cones moves along the path $\vec{M}-\vec{\Gamma}$, while the other moves along the $\vec{M}-\vec{K}/\vec{K}'$ direction, being the latter responsible for the formation of the cone at $\vec{K}$ point, while the former forms the parabolic band crossing point at $\vec{\Gamma}$ point in the Kagome lattice \cite{tony2019, PhysRevA.107.023509}. Similar to the Lieb lattice, two smaller peaks associated with the van Hove singularities are observed in the density of states [see Fig.~\ref{fEspectro_Transicao_mon}(a)]. However, unlike the pronounced single-peak caused by the flat band in the Lieb lattice, now one has an additional second peak around $E/t \approx 0$ with a slightly smaller amplitude. In the 3D spectrum, Fig.~\ref{fEspectro_Transicao_mon}(b), it is observed that for each Dirac cone, there is the formation of two tilted cones, one pair originating from the connection with the upper and middle bands, and another pair from the lower and middle bands, totaling eight tilted cones.

\subsection{Kagome lattice}\label{sec.mono.kagome}

\begin{figure}[h!]
    \centering
    \includegraphics[scale=0.2]{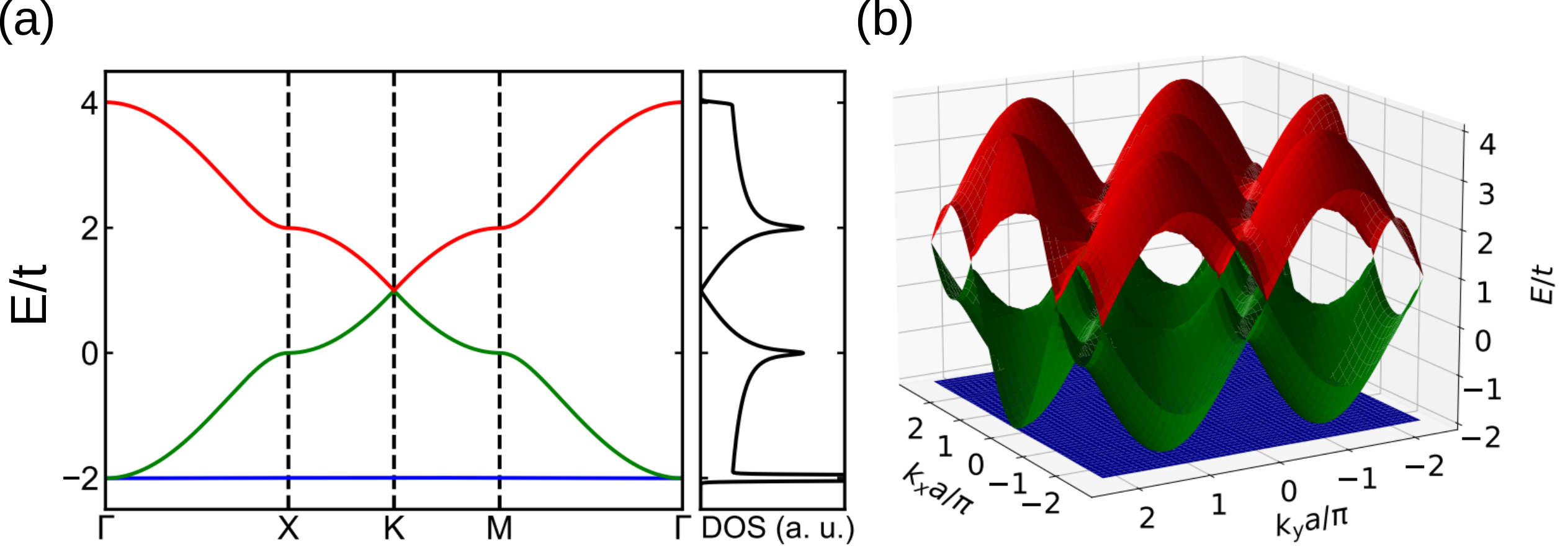}
    \caption{The same as in Fig.~\ref{Espectro_lieb}, but now for the Kagome lattice ($\theta = 2\pi/3$). The path connecting the high-symmetry points is now $\vec{\Gamma}-\vec{X}-\vec{K}-\vec{M}-\vec{\Gamma}$.}
    \label{Espectro_kagome_mon}
\end{figure}

Figure~\ref{Espectro_kagome_mon} depicts the band structures for the Kagome lattice in (a) the 2D format along the high-symmetry points and (b) in 3D format, obtained by diagonalizing the tight-binding Hamiltonian \eqref{eqh5} with $\theta = 3\pi/2$. Three energy bands are observed: two dispersive ones, in which they form the Dirac cone at $\vec{K}$ point, and a flat band \cite{PhysRevB.80.193304, PhysRevB.80.113102, mojarro2023strain}. Similarly to Lieb and transition lattices, the density of states for the Kagome lattice exhibits two peaks associated with the van-Hove singularities and a pronounced peak characteristic of the flat band. Once we assumed $t<0$, the flat band is energetically positioned at the bottom of the band structure at $E/t=-2$. \cite{PhysRevB.108.125433, wellissonTPT}

\section{Multilayer Lieb-Kagome lattice}\label{sec.II}

In this section, we employ the tight-binding model to derive the theoretical formalism for treating a multilayer generic Lieb-Kagome lattice. We first construct in Sec.~\ref{sec.bilayer} the tight-binding Hamiltonian for the bilayer case, considering the two most common types of stacking: AA and AB (Bernal). Thereafter, in Sec.~\ref{sec.multilayer}, the model is generalized to describe the electronic properties of the multilayer system, considering the coupling solely between adjacent layers.

\subsection{Bilayer}\label{sec.bilayer}

\begin{figure}[t]
    \centering
    \includegraphics[scale=0.2]{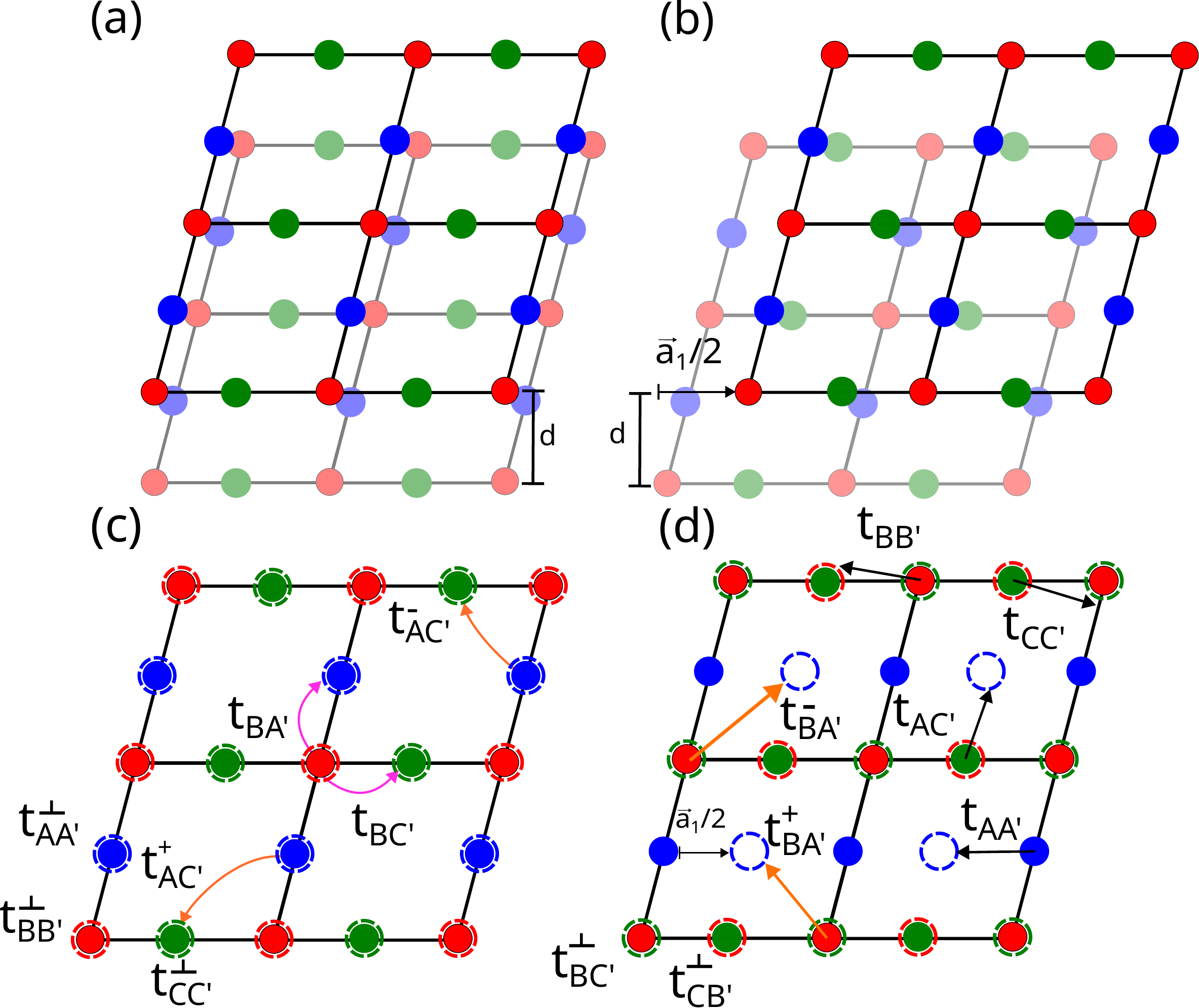}
    \caption{Schematic illustrations of the bilayer transition lattice ($\theta = 7\pi/12$) in a perspective view, assuming the (a) AA and (b) AB stacking. A top view is provided for the (c) AA and (d) AB stacking, highlighting the interlayer hopping configurations considered in the model. The filled circles represent the bottom layer sites, while the dashed circles refer to the top layer sites. The hopping parameters $t_{ij'}^{\perp}$ and $t_{ij'}$ indicate the perpendicular and the diagonal hoppings, respectively, with $i$ referring to the $i$-th bottom layer sublattice and $j'$ to the $j$-th top layer sublattice.}
    \label{equema_bicamada}
\end{figure}

The generic bilayer system is formed by stacking two monolayers of the Lieb-Kagome lattice, \textit{i.e.} it consists of two coupled generic lattices by the interlayer hopping parameters, possessing six nonequivalent sites per unit cell, being A, B, C (A', B', C') on the bottom (top) layer. Sketches of the generic bilayer system, taking the transition lattice as an example, are presented in Fig.~\ref{equema_bicamada}, assuming the two most common stacks, AA [Figs.~\ref{equema_bicamada}(a) and \ref{equema_bicamada}(c)] and AB [Figs.~\ref{equema_bicamada}(b) and \ref{equema_bicamada}(d)]. In AA-stacking [Figs.~\ref{equema_bicamada}(a) and \ref{equema_bicamada}(c)], the two single layers are aligned, such that the sites in the top layer (${A}', {B}', {C}'$) are located directly above the sites in the bottom layer (${A}, {B}, {C}$). In the AB-stacked case [Figs.~\ref{equema_bicamada}(b) and \ref{equema_bicamada}(d)], the top layer is shifted by $\vec{a}_1/2$ along the horizontal direction with relation to the bottom layer.

Since the unit cell of the generic bilayer system contains six non-equivalent sites (A, B, C, A', B', C'), the tight-binding Hamiltonian can be written as
\begin{equation}\label{eqh7}
    \hat{\hham}_{bi} = \sum_{i=1}^2 \hat{\hham}_{sl}^{(i)} + \hat{\hham}_{int},
\end{equation}
where $\hat{\hham}_{sl}^{(i)}$ is the Hamiltonian describind each single layer $i$, and $\hat{\hham}_{int}$ being
\begin{align}
        \hat{\hham}_{int} &= \sum_{ij}\left [ t_{A{A}'} \left ( {a}'^{\dagger}_{i}{a}_{j} + {a}^{\dagger}_{i}{a}'_{j}\right ) +t_{A{B}'} \left ( {b}'^{\dagger}_{i}{a}_{j}+ {a}^{\dagger}_{i}{b}'_{j}\right ) \right. \nonumber \\ 
        & \left. +t_{A{C}'} \left ( {c}'^{\dagger}_{i}{a}_{j} + {a}^{\dagger}_{i}{c}'_{j}\right )\right ] \nonumber \\ 
        &+\sum_{ij}\left [ t_{B{A}'} \left ( {a}'^{\dagger}_{i}{b}_{j} + {b}^{\dagger}_{i}{a}'_{j}\right ) +t_{B{B}'} \left ( {b}'^{\dagger}_{i}{b}_{j}+ {b}^{\dagger}_{i}{b}'_{j}\right ) \right. \nonumber \\ 
        & \left. +t_{B{C}'} \left ( {c}'^{\dagger}_{i}{b}_{j} + {b}^{\dagger}_{i}{c}'_{j}\right )\right ] \nonumber \\ 
        &+\sum_{ij} \left[ t_{C{A}'}  \left (  {a}'^{\dagger}_{i}{c}_{j} + {c}^{\dagger}_{i}{a}'_{j}  \right) +  t_{C{B}'}  \left ( {b}'^{\dagger}_{i}{c}_{j} + {c}^{\dagger}_{i}{b}'_{j}  \right ) \right. \nonumber \\ 
        & \left. +t_{C{C}'} \left ( {c}'^{\dagger}_{i}{c}_{j}  +  {c}^{\dagger}_{i}{c}'_{j} \right ) \right ],
\end{align}
where $t_{ij}$ is the interlayer hopping parameter, \textit{i.e.} energetic connections between the sites of the bottom layer ($i = A, B$, and $C$) and the sites of the top layer ($j = A', B'$, and $C'$).

Concerning the range of interlayer hopping parameters, we consider the amplitudes to be finite for first nearest neighbors ${t}_{i{j}}^{\perp}$ located on $\vec{a}_{3}$, the second nearest neighbors ${t}_{i{j}}$ on $\pm\vec{a}_1/2 + \vec{a}_{3 }$ and $\pm\vec{a}_2/2 + \vec{a}_{3}$, and the third nearest neighbors ${t}_{i{j}}^{+(-)}$ on $\pm\left(\vec{a}_1-\vec{a}_2\right)/2 + \vec{a}_{3}$ and $\pm\left(\vec{a}_1+\vec{a}_2\right)/2 + \vec{a}_{3}$, where $\vec{a}_{3} = d\hat{\upsilon}_{3}$ with $\hat{\upsilon}_{3} = (0,0,1)$ and $d$ being the distance between the layers, assuming the same interlayer distance $d$ for both AA and AB stacks, changing only the energetic connections between the lower and upper layer sites. For instance, in the AA-stacked case, each site ${i}'$ of the upper layer is linked directly to a site ${i}$ of the bottom layer, in which the interlayer hopping is ${t}_{i{i}'}^{\perp}$ with $i = \{ A, B, C\}$, whereas for the AB-stacked case, the interlayer nearest-neighbor hopping is due to sites $B \leftrightarrow {C}'$ and $C \leftrightarrow {B}'$, given by ${t}_{B{C}'}^{\perp}$ and ${t}_{C{B}'}^{\perp}$, respectively. Figures~\ref{equema_bicamada}(c) and \ref{equema_bicamada}(d) show the interlayer hopping parameters for AA and AB stacks, respectively.

By Fourier transforming the field operators in Eq.~\eqref{eqh7}, using Eq.~\eqref{fourie}, and repeating the theoretical procedure employed in the previous section, one ends up with the bilayer Hamiltonian and eigenstates written in momentum space, such as
\begin{equation}\label{eqh8}
 \hat{\hham}_{bi}(\vec{k}) = \begin{pmatrix}
\hat{\hham}_{\vec{k}} & \hat{\hham}_{C} \\ 
 \hat{\hham}_{C}^{\dagger}& \hat{\hham}_{\vec{k}}
\end{pmatrix},\;\;\;\textup{and}\;\;\; \Phi_{\vec{k}} = 
\begin{pmatrix}
\psi_{\vec{k}}^{(1)}\\ 
\psi_{\vec{k}}^{(2)}
\end{pmatrix},
\end{equation}
with $\psi_{\vec{k}}^{(i)} =  \left(\phi_{a ({a}')}\;\;\;\;  \phi_{b({b}')}\;\;\;\;  \phi_{c({c}')}\right)^{T} $, where $i = \{1,2\}$ is the layer index. The bilayer Hamiltonian $\hat{\hham}_{bi}(\vec{k})$ [Eq.~\eqref{eqh8}] is a $6\times6$ matrix that can be written in a $2\times 2$ block matrix format in which the main diagonal blocks ($\hat{\hham}_{\vec{k}}$) correspond to the monolayer Hamiltonian [Eq.~\eqref{eqh5}], and the off-diagonal blocks ($\hat{\hham}_{C}$) correspond to the interaction between adjacent layers, given by
\begin{equation}\label{eqh9}
    \hat{\hham}_{C} = \begin{pmatrix}
\hham_{A{A}'} & \hham_{A{B}'}  & \hham_{A{C}'} \\ 
\hham_{B{A}'}& \hham_{B{B}'}  & \hham_{B{C}'} \\ 
\hham_{C{A}'} &H_{C{B}'}  &\hham_{C{C}'} 
\end{pmatrix}.
\end{equation}
According to Eq.~\eqref{hopping1}, the position-dependent hopping parameters are real function, such that $t_{ij}^* = t_{ij}$; therefore, all matrix elements of $\hat{\hham}_{C}$ are also real functions, leading to $\hham_{A{B}'} = \hham_{B{A}'}$, $\hham_{A{C }'} = \hham_{C{A}'}$, and $\hham_{B{C}'} = \hham_{C{B}'}$. 

For the AA-stacking case, the matrix elements of $\hat{\hham}_{C}$ are given by
\begin{subequations}\label{eqh10}
	\begin{align}
	\hham_{j{j}'}(\vec{k}) &= t_{j{j}'}^{\perp}\;\;(j = A,\; B,\; C),\\    \hham_{B{A}'}(\vec{k})&=2t_{B{A}'}\cos\left(a_0\vec{k}\cdot\hat{\upsilon}_2\right),\\
	\hham_{B{C}'}(\vec{k})&=2t_{B{C}'}\cos\left(a_0\vec{k}\cdot\hat{\upsilon}_1\right),\\
	\hham_{A{C}'}(\vec{k})&=2{t_{A{C}'}^{-}}\cos\left[a_0\vec{k}\cdot(\hat{\upsilon}_1-\hat{\upsilon}_2)\right]\nonumber\\
&+2{t_{A{C}'}^{+}}\cos\left[a_0\vec{k}\cdot(\hat{\upsilon}_1+\hat{\upsilon}_2)\right],
	\end{align}
\end{subequations}
with $\hat{\upsilon}_1=(1, 0,0)$ and $\hat{\upsilon}_2=\left(-\cos\theta,\sin\theta, 0\right)$. Comparing the matrix elements of Eq.~\eqref{eqh10} with those for the monolayer case given by Eq.~\eqref{eqh6}, one clearly notices the similarity between the AA stacking interactions and the monolayer Hamiltonian terms.

For the AB-stacking case, the top layer is shifted by ($\vec{a}_1/2$) horizontally. Due to this position displacement of the top layer sites with respect to the bottom layer sites, the interlayer interactions change, and thus the matrix elements of the interaction Hamiltonian $\hat{\hham}_{C}$ become
\begin{subequations}\label{eqh11}
	\begin{align}
	\hham_{j{j}'}(\vec{k}) &= 2t_{j{j}'}^{\perp}\cos\left(a_0\vec{k}\cdot\hat{\upsilon}_1\right)\;\;(j = A,\; B,\; C),\\    
\hham_{B{A}'}(\vec{k})&=2{t_{B{A}'}^{-}}\cos\left[a_0\vec{k}\cdot(\hat{\upsilon}_1-\hat{\upsilon}_2)\right]\nonumber\\
&+2{t_{B{A}'}^{+}}\cos\left[a_0\vec{k}\cdot(\hat{\upsilon}_1+\hat{\upsilon}_2)\right],\\
	\hham_{B{C}'}(\vec{k})&=t_{B{C}'}^{\perp},\\
	\hham_{A{C}'}(\vec{k})&=2{t_{A{C}'}}\cos\left(a_0\vec{k}\cdot\hat{\upsilon}_2\right).
	\end{align}
\end{subequations}
Replacing the aforementioned expressions, Eq.~\eqref{eqh10} for the AA-stacking and Eq.~\eqref{eqh11} for the AB-stacking, into the interaction Hamiltonian $\hat{\hham}_{C}$, and this, in turn, into Eq.~\eqref{eqh8} of the bilayer Hamiltonian, one can obtain the homobilayer Lieb-Kagome band structures by diagonalizing $\hat{\hham}_{bi}$.

\subsection{Multilayer}\label{sec.multilayer}

The bilayer Lieb-Kagome Hamiltonian [Eq.~\eqref{eqh8}] can be easily generalized to the multilayer system case formed by the generic Lieb-Kagome lattice with either AA or AB stacks. The Hamiltonian for an $N$-layer system can be expressed by
\begin{equation}\label{eqh13}
        \hat{\hham}_{N} = \begin{pmatrix}
\hat{\hham}_{\vec{k}} & \hat{\hham}_C &  &  &  & \\ 
 \hat{\hham}_{C}^{\dagger}& \hat{\hham}_{\vec{k}} & \hat{\hham}_C &  &  & \\ 
 &\hat{\hham}_{C}^{\dagger}& \hat{\hham}_{\Vec{k}}  & \hat{\hham}_{C}  & &  & \\ 
 &  &  & \ddots  &  & \\ 
 &  &  &  &  & \hat{\hham}_{C}\\ 
 &  &  &  &\hat{\hham}_{C}^{\dagger}  & \hat{\hham}_{\vec{k}} 
\end{pmatrix}_{N\times N},
\end{equation}
with the eigenstates given by
\begin{equation}\label{eqh14}
\Phi_{N} = \begin{pmatrix}
\psi_{\vec{k}}^{(1)}\\ 
\psi_{\vec{k}}^{(2)}\\ 
\psi_{\vec{k}}^{(3)}\\ 
\vdots \\ 
\psi_{\vec{k}}^{(N)}
\end{pmatrix}_{N\times1},    
\end{equation}
with $\psi_{\vec{k}}^{(i)} = \left ( \phi_{a,i}\;\;\;\phi_{b,i}\;\;\;\phi_{ c,i} \right )^{T}$, being $i$ the layer index. Note that the Hamiltonian $\hat{\hham}_{N}$ [Eq.~\eqref{eqh13}] is a tridiagonal matrix formed by $3\times 3$ blocks since only interactions between adjacent layers are considered. The main diagonal elements ($\hat{\hham}_{\vec{k}}$) correspond to the monolayer Hamiltonian [Eq.~\eqref{eqh5}], and the adjacent diagonals are composed of the interaction matrix ($\hat{\hham}_{C}$) [Eq.~\eqref{eqh9}], which will depend on the type of stacking considered.

\begin{figure*}[t]
{\includegraphics[width=0.8\linewidth]{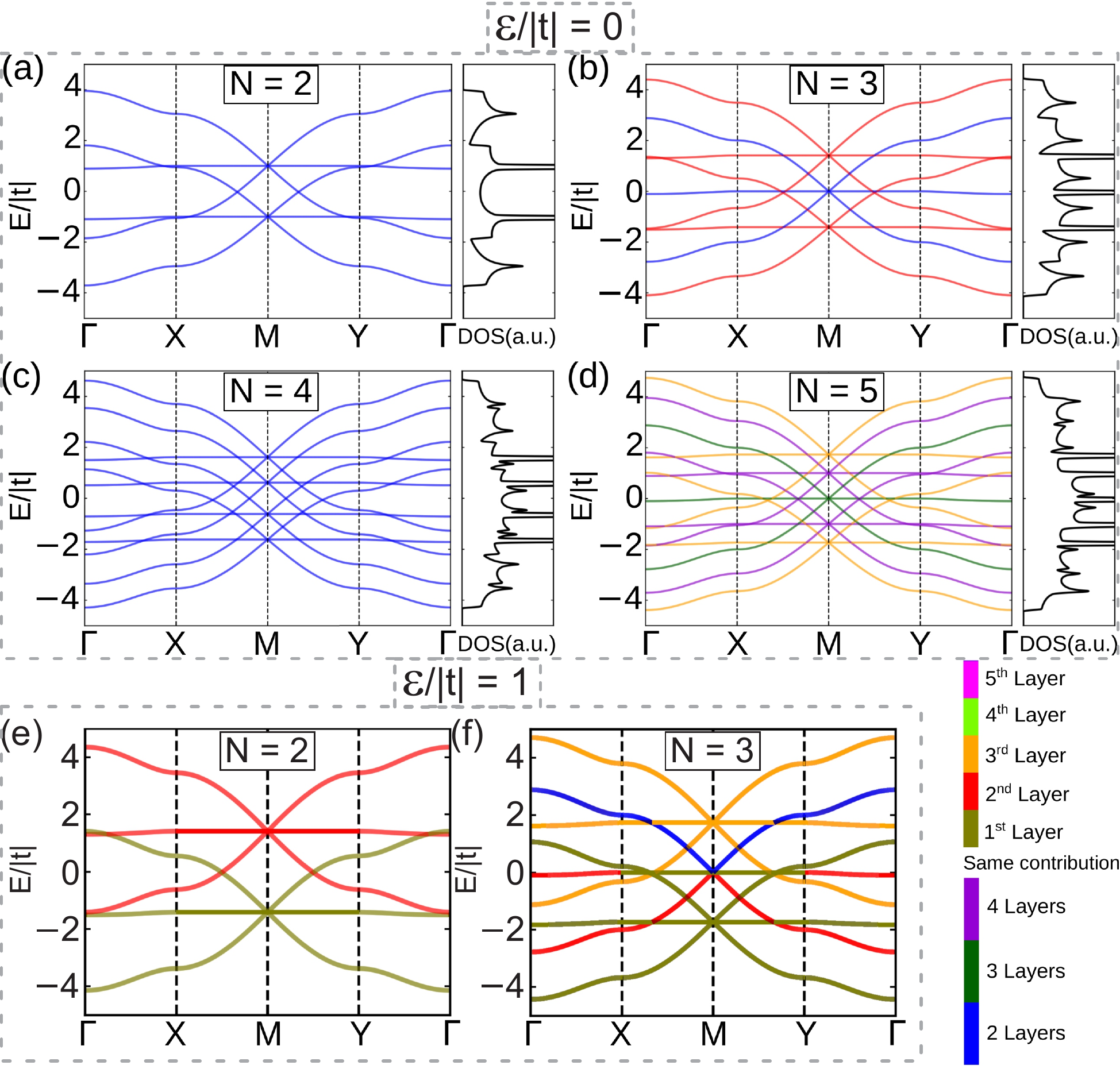}}
\caption{Multilayer Lieb lattice band structures, projected onto the layer with the highest contribution $\left | \braket{\psi_{\vec{k}}^{i}|\Phi} \right |^{2}$, assuming different numbers of layers: (a) $N = 2$, (b) $N = 3$, (c) $N = 4$, and (d) $N = 5$. The color scheme indicates the layer with the highest projected probability density's contribution (brown, red, yellow, light-green, and pink colors denote the highest contribution at the $1^\text{st}$, $2^\text{nd}$, $3^\text{rd}$, $4^\text{th}$, and $5^\text{th}$ layer, respectively) or the number of layers with the same contribution (blue, olive, and purple colors denote equal contributions of two, three, and four layers, respectively). The right panels of (a)-(d) show the corresponding density of states. Lieb lattice electronic spectra for the (e) bilayer and (f) trilayer cases are also depicted when a potential difference equal to $\epsilon/|t| = 1$ is applied.}
    \label{espectro_lib_mult}
\end{figure*}

\section{AA-stacking}\label{sec.III}

In this section, the electronic properties of the multilayer system are discussed, considering the AA stacking for the Lieb $\left(\theta = \pi / 2\right)$, transition $\left(\pi/2 < \theta < 2\pi/3\right)$, and Kagome $\left(\theta = 2\pi/3\right)$ lattices, respectively, in order to assess how the number of layers and an electrostatic bias tune the electronic characteristics of these lattices. The AA-stacked Lieb-Kagome multilayer system consists of $3N$ sites per unit cell, in which each layer is aligned with the adjacent layers, such that the sites $a_i$, $b_i$, and $c_i$ of layer $i$ are perpendicularly connected with the sites $a_{i\pm 1}$, $b_{i\pm 1}$, and $c_{i\pm 1}$ of the adjacent layers on the top ($i+ 1$) and on the bottom ($i - 1$). As a result, the band structure has $3N$ energy bands.

\subsection{Lieb lattice}\label{BI_Lieb_lattice}

The band structure for the $N$-layered Lieb lattice is obtained by diagonalizing the multilayer Hamiltonian [Eq.~\eqref{eqh13}], taking $\theta = \pi /2$ into the matrix elements of the interaction Hamiltonian described by Eq.~\eqref{eqh10}. Figure~\ref{espectro_lib_mult} presents the energy spectra and the densities of states of multilayer Lieb lattice considering (a) $N=2$, (b) $N=3$, (c) $N=4$, and (d) $N=5$ layers. For the bilayer Lieb case ($N=2$), as displayed in Fig.~\ref{espectro_lib_mult}(a), one can observe two sets of monolayer Lieb energy bands, energetically shifted with respect to each other, where each band's set is composed of three energy bands originating from the monolayer energy spectrum. Henceforth, we shall refer to the energetic set of three monolayer Lieb energy bands as Lieb Bands (LB).

As the number of layers increases, new LB sets emerge in the energy spectrum, all shifted in energy, as evidenced in Figs.~\ref{espectro_lib_mult}(b)-\ref{espectro_lib_mult}(d). The presence of these additional bands indicates the cumulative influence of the layers with an extra LB set per layer on the band structure of the system. Such a statement can also be verified in the corresponding densities of states on the right side of each band structure in Fig.~\ref{espectro_lib_mult}. As in the monolayer case [Fig.~\ref{Espectro_lieb}(a)], the density of states contains characteristic peaks of the flat bands and several smaller peaks associated with the van-Hove singularities. This cumulative feature in the band structure and the density of states is a characteristic of systems with the AA stacking configuration. \cite{ROZHKOV20161, doi:10.1143/JPSJ.76.024701, PhysRevB.96.115426}.

To understand the reason why the energy spectrum of a Lieb lattice $N$-layer system consists of a copy of $N$ monolayer energy bands shifted in energy, let us take the bilayer case as an example. In the AA stacking, the reflection plane between the layers is a symmetry operation $\hat{\sigma}_{h}$ that keeps the lattice invariant, that is, it commutes with the Hamiltonian, $\left [\hat{\hham}_{bi}, \hat{\sigma}_{h} \right] = 0$. Thus, one can use the eigenstates of the reflection operator $\hat{\sigma}_{h}$ to rewrite the bilayer Hamiltonian through a unitary transformation, $\hat{P}^{-1}\hat{\hham}_{bi} \hat{P}$, where $\hat{P}$ is the matrix composed of the eigenstates of the reflection operator. In the layer space, the reflection operator can be written as $\hat{\sigma}_{h} = \tau_{x}$, with $\tau_{x}$ being a Pauli matrix in the layer space. From the eigenvalues of the reflection operator, $m = \pm 1$, one observes that there are symmetric states ($m = 1$) and antisymmetric states ($m = -1$) for the reflection plane. Diagonalizing the Hamiltonian after the unitary transformation, one obtains
\begin{equation}\label{auto1}
E^{\pm}_{n} = \varepsilon_{n}(\vec{k}) \pm \left | t^{\perp}\right|,   
\end{equation}
with the eigenvectors being given by
\begin{equation}\label{auto2}
    \Psi^{\pm}_{n,k} = \frac{1}{\sqrt{2}} \left(\psi_{n,\vec{k}}^{(1)}  \pm \psi_{n,\vec{k}}^{(2)}\right),
\end{equation}
being a $3\times 1$ eigenvector for the bonding and anti-bonding states, and $\varepsilon_{n}(\vec{k})$ corresponding to the eigenvalues of the monolayer Lieb case. Therefore, it is evident from Eq.~\eqref{auto1} that the bilayer band structure consists of two copies of the monolayer energy bands, in which one of the LB is energetically shifted up by $+\left|t^{\perp}\right|$ and the other LB is shifted down by $-\left|t^{\perp}\right|$, leading to an energy separation between these two bands' sets of $E^{+} - E^{-} = 2\left|t^{\perp}\right|$. If the coupling between the layers is zero, $t^{\perp} = 0$, the bilayer band structure will correspond to a doubly degenerate monolayer Lieb band structure. A discussion on the role played by the hopping energies between sites in adjacent layers with distances greater than the interlayer distance ($a_{ij}>d$), \textit{i.e.}, when one takes or not interlayer connections beyond the first neighbors, is given in Sec.~\textcolor{blue}{SIII} of the Supplemental Material \cite{SI}. As shown in Figs.~\textcolor{blue}{S23}(a) and \textcolor{blue}{S23}(d), a comparison between the band structures for the bilayer and trilayer Lieb lattice is presented, respectively, for cases with ($t^\perp + t_{ij}^\perp$ - blue curves) and without ($t^\perp$ - red curves) diagonal interlayer hoppings, demonstrating that second-neighbor interlayer hoppings cause solely a slight energy shift in the bands. This can also be verified in Fig.~\textcolor{blue}{S24}(a) for the density of states of the bilayer Lieb lattice. Thus, one can state that a first-neighbor tight-binding model for interlayer connections would be theoretically sufficient to capture the physics of such AA-stacked multilayer flat-band systems.

The spatial distribution of the electronic states in a material is a crucial characteristic for understanding and controlling its physical properties \cite{PhysRevB.96.115426, crasto2019layertronic, jaskolski2018controlling, morozov2005two}. In this context, it is essential to know how each layer contributes to a given state, $\left | \braket{\psi_{\vec{k}}^{(i)}|\Phi} \right |^{2}$. To perform this study, we evaluate the contribution of each layer at each point $\vec{k}$ where the band structure is calculated. We then label the state using a color scheme to indicate the layer representing the most significant contribution. Figure~\ref{espectro_lib_mult} shows the AA-stacked multilayer Lieb lattice band structures projected onto the layer with the highest contribution, \textit{i.e.}, the layer with the highest probability of finding the electron. One notices for the bilayer case [Fig.~\ref{espectro_lib_mult}(a)] that the two layers contribute equally (blue color) to all states, due to the reflection symmetry between the layers. This can be seen from the eigenvectors of the Hamiltonian, given by Eq.~\eqref{auto2}, which, due to the existence of reflection symmetry one has that $\left | \braket{\psi_{n,\vec{k}}^{(i)}|\Psi^{\pm}_{n,k}} \right|^{2} = 1/2$.

For the trilayer Lieb case, Figure~\ref{espectro_lib_mult}(b) shows that electronic states associated with the highest and lowest energetic LBs have their probability distributions preferably localized in the middle layer ($2$nd layer - red curves), whereas the electronic states associated with the middle LBs (blue curves) have their probability distributions equally distributed in two layers: the $1$st and the $3$rd layers. For the tetralayer case ($N = 4$), the projected band structure in Fig.~\ref{espectro_lib_mult}(c) has the probability distributions of the electronic states for all the twelve bands preferably localized in two layers (blue curves). From the spatial distribution along the $z$-direction of the wave functions, we verified (although not shown here) that the states are symmetrically distributed, being either more localized in the two middle layers (\textit{i.e.}, $2$nd and $3$rd layers) or in the outer layers (\textit{i.e.}, $1$st and $4$th layers). In the case of $N = 5$ [Fig.~\ref{espectro_lib_mult}(d)], one observes that the probability distributions for the lowest and highest energy LBs (orange bands) are more localized in the $3$rd layer, whereas the set of LBs around $E/|t|=0$ (green curves) have their wave functions preferably distributed symmetrically in three layers, namely the $1$st, $3$rd, and $5$th layers. The second and fourth sets of LBs, labeled in purple, have their states distributed across four layers, located in the $1^\text{st}$, $2^\text{nd}$, $4^\text{th}$, and $5^\text{th}$ layers. Through this analysis, it is evident that the spatial localization of the probability distributions projected per band with respect to the number of layers exhibits a sinusoidal behavior. This occurs because, with the increase in the number of layers, the Hamiltonian of the multilayer system in the stacking direction ($z$-axis) becomes equivalent to the tight-binding Hamiltonian of a one-dimensional chain with $N$ sites, like the simple Kronig-Penney model. \cite{grosso2013solid} In this simple quantum picture, one can see the $N$-layer Lieb lattice along the $z$-direction as: (i) a periodic superlattice with $N$ quantum wells or even as (ii) an effective infinite well of size $L = (N+1)d$, where $d$ is the distance between the layers. Following such analogies, the electronic probability distributions along the $z$-direction in a multilayer Lieb lattice have a nodal pattern similar to that of the confined states in an infinite well or a 1D Kronig-Penney-like periodic potential \cite{grosso2013solid, crasto2019layertronic}. To check that, note that for $N = 5$, the set of LBs in orange would correspond to the ground state in the analog aforementioned systems with a peak in the central layer, whereas the set of LBs in green (purple) would correspond to the second (third) excited state with two (third) peaks.

In the context of state localization, it is expected that applying an external electric field will cause the breaking of reflection symmetry between the layers, thus altering the spatial contribution of each layer in the electronic probability distribution. This change in layer-dependent spatial contribution modulates the system's properties, providing an effective tool for tuning the material's electronic properties. The perpendicular electric field is included by adding a potential difference between the layers. For the bilayer case, the total Hamiltonian reads now
\begin{equation}\label{campo_1}
    \hat{\hham}_{T} = \hat{\hham}_{bi} - \epsilon\mathbb{I}_{3\times 3}\otimes\tau_{z},
\end{equation}
where $\epsilon$ is the bias voltage amplitude, $\tau_{z}$ is the $z$ Pauli matrix, and the Kronecker tensor product $\mathbb{I}_{3x3}\otimes\tau_{z}$ combines the $3\times 3$ identity ($\mathbb{I}_{3\times 3}$) with the $2 \times 2$ Pauli matrix yielding a $6\times 6$ matricial bias-related term in the total Hamiltonian \eqref{campo_1}. Figures~\ref{espectro_lib_mult}(a) and \ref{espectro_lib_mult}(e) and Figs.~\ref{espectro_lib_mult}(b) and \ref{espectro_lib_mult}(f) show the bilayer and trilayer band structures, respectively, with a bias gate of (a, b) $\epsilon/|t| = 0$ and (e, f) $\epsilon/|t| = 1$. When a perpendicular electric field is applied ($\epsilon/|t| \neq 0$) to the multilayer Lieb system, no pronounced qualitative changes in the band structure are observed, \cite{ROZHKOV20161, doi:10.1143/JPSJ.76.024701} except by the imbalance of the wave functions' contributions per layer and an increase in the energy separation between the sets of LBs \cite{PhysRevB.96.115426, crasto2019layertronic}. This imbalance of the wave function distribution can be understood in terms of the eigenstates of the Hamiltonian operator. Since the reflection symmetry between the layers is broken due to the application of the electric field, one can write an effective reflection operator such as $ \widetilde{\hat{\sigma}_{h}} = \left ( t^{\perp}\tau_{x}-\epsilon\tau_{z} \right )/\sqrt{\epsilon^{2} + (t^{\perp})^{2}}$, whose operator construction followed the same approach demonstrated in the Supplementary Material of Ref.~[\onlinecite{PhysRevB.96.115426}]. Using the eigenvectors of the operator $ \widetilde{\hat{\sigma}_{h}}$ and applying the similarity transformation to the Hamiltonian \eqref{campo_1}, one finds that the eigenvalues are given by
\begin{equation}\label{campo_2}
E_{n}^{\pm} = \varepsilon_{n}(\vec{k}) \pm \sqrt{\epsilon^{2} + (t^{\perp})^{2}},    
\end{equation}
where $\varepsilon_{n}(\vec{k})$ corresponds to the eigenvalues for the monolayer Lieb lattice, $t^{\perp}$ being the perpendicular interlayer hopping, and $\epsilon$ is the bias potential. Thus, the energetic separation between each set of LBs depends on the electrostatic potential in each layer, that is, $E^{+}_{n} - E^{-}_{n} = 2\sqrt{\epsilon ^{2} + (t^{\perp})^{2}}$. The eigenvectors of the bilayer Lieb Hamiltonian are given by the following bonding $\left(\Psi_{n,\vec{k}}^{+}\right)$ and anti-bonding $\left(\Psi_{n,\vec{k}}^{-}\right)$ states
\begin{equation}\label{campo_3}
    \Psi_{n,\vec{k}}^{\pm} = C_{\pm}\psi_{\vec{k}}^{(1)}
    \pm D_{\pm}\psi_{\vec{k}}^{(2)},
\end{equation}
with $C_{\pm}$ and $D_{\pm}$ being 
\begin{equation}\label{campo_4}
    C_{\pm} = \left (1 + D_{\pm}^{2} \right)^{-1/2}\;\;\;\mbox{and}\;\;\; D_{\pm} = \frac{\left [ \epsilon \pm \sqrt{\epsilon^{2} + (t^{\perp})^{2}}\right]}{t^{\perp}},
\end{equation}
and $\Psi_{n,\vec{k}}^{\pm}$ are $3\times 1$ matrices. Therefore, from the bias dependence $\epsilon$ on Eqs.~\eqref{campo_3} and \eqref{campo_4}, one clearly notes that the probability distribution per layer for a given state can be modulated by the applied perpenficular electric field, being $\left|C_{\pm}\right|^{2}$ and $1 - \left|C_{\pm}\right|^{2}$ associated with the contributions on the first and second layers, respectively. Such imbalance of the wave function's contributions can be observed in Figs.~\ref{espectro_lib_mult}(e) and \ref{espectro_lib_mult}(f) for $N=2$ and $N=3$, in which one notices a tendency for the states to be more localized in the outermost layers, following the energy displacement caused by the electric field. For instance, for $N=3$ in Fig.~\ref{espectro_lib_mult}(f), see that the upper and lower sets of LBs in yellow and moss green colors correspond to energy bands in which the spatial localizations of the probability distribution are preferably in the third and first layers, respectively. By the color label in Fig.~\ref{espectro_lib_mult}(f), note that the electronic states associated with the central LBs are partially distributed in different layers, with a tendency to be more localized in the central layer ($2$nd layer). For the bilayer case [Fig.~\ref{espectro_lib_mult}(e)], the high-energetic set of energy bands is more localized in the upper layer (red curves - $2$nd layer), whereas the low-energy bands have their electronic states distributed preferentially in the lower layer (moss green curves - $1$st layer). 

\begin{figure*}[t]
    \centering
    \includegraphics[width=0.8\linewidth]{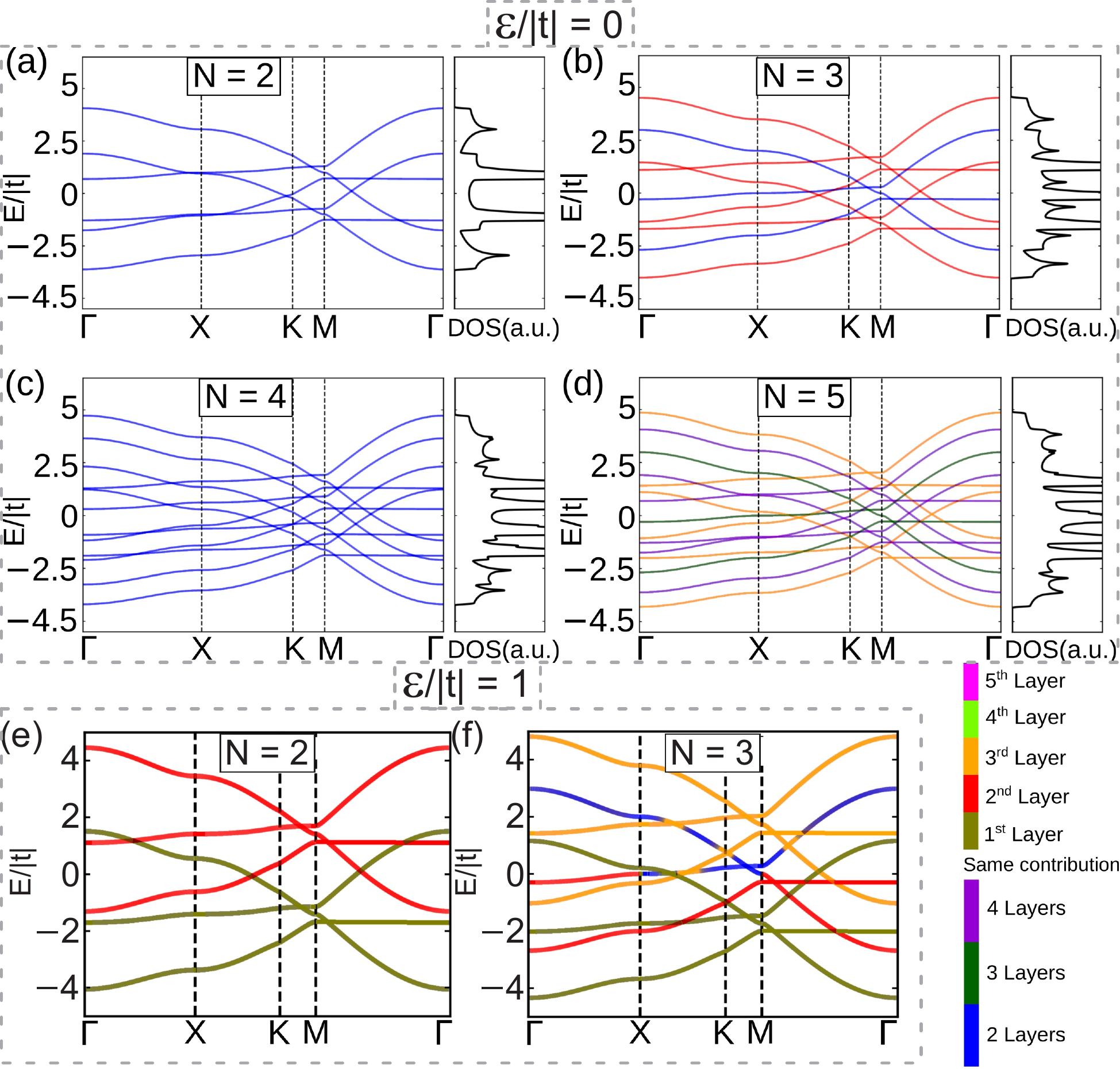}
    \caption{The same as in Fig.~\ref{espectro_lib_mult}, but now for AA-stacked multilayer transition lattice, assuming $\theta = 7\pi/12$.}
    \label{espectro_transicao_mult}
\end{figure*}

To better understand the $z$-distribution and layer-dependence of the electronic states in the multilayer Lieb system, Fig.~\textcolor{blue}{S2} in the Supplemental Material \cite{SI} shows the band structure of the bilayer Lieb lattice projected onto the first (top row of panels) and second layer (bottom row of panels) for three potential differences, highlighting the field's effect on the state distribution, and Figs.~\textcolor{blue}{S3}(a) and \textcolor{blue}{S3}(b) present the density of states for the bilayer and trilayer Lieb lattices for three potential differences, respectively. From the density of states in Fig.~\textcolor{blue}{S3}, one observes a shift in the position of the van Hove singularities due to the field-induced separation in the bilayer energetic bands. The perpendicular electric field increases the separation between energy bands, with high-intensity fields completely separating the two sets of LBs, for instance, see Figs.~\textcolor{blue}{S2}(c) and \textcolor{blue}{S2}(f) for $\epsilon/|t|=1$ such that exhibits a significant wave function distribution imbalance and a pronounced energetic displacement between the two sets of LBs. This band energetic separation effect can also be achieved by adjusting the interlayer distance, which is equivalent to modifying the intensity of the perpendicular hopping ($t^\perp$). To verify that, we show in Fig.~\textcolor{blue}{S4} the AA-stacked bilayer and trilayer band structures for different interlayer distances $d/a_0$ and the corresponding wave functions. For large distances, the energy spectra become analogous to decoupled two- and three-layer Lieb systems, \textit{i.e.} with doubly and triply degenerate LBs for $N=2$ and $N=3$, respectively [see Figs.~\textcolor{blue}{S4}(a) and \textcolor{blue}{S4}(d)]. For smaller distances, the energetic separation between the LBs increases [Fig.~\textcolor{blue}{S4}(c)], and the probability densities of the electronic states are mainly distributed in the outermost layers [Fig.~\textcolor{blue}{S4}(b)]. Additional discussions on the effect of a perpendicular electric field on the AA-stacked multilayer Lieb lattice band structure and its density of states, concerning Figs.~\textcolor{blue}{S2}--\textcolor{blue}{S4}, are presented in Sec.~\textcolor{blue}{SIIA1} in the Supplemental Material \cite{SI}.

\subsection{Transition lattice}\label{BI_Rede_Transição}

The band structure, density of states, and the corresponding probability densities for the AA-stacked $N$-layer transition system are discussed in this section, being then obtained by diagonalizing the multilayer Hamiltonian [Eq.~\eqref{eqh13}], taking $\theta = 7\pi / 12$, as an example, into the matrix elements of the interaction Hamiltonian described by Eq.~\eqref{eqh10}. Figure~\ref{espectro_transicao_mult} shows the AA-stacked multilayer transition lattice band structures considering (a) $N=2$, (b) $N=3$, (c) $N=4$, and (d) $N=5$ layers in the absence of any external field. For $N=2$, Fig.~\ref{espectro_transicao_mult}(a) shows two equal sets of monolayer transition bands, labeled from now on as transition band (TB), being one of the energetic sets of TB shifted upward, whereas the other is shifted downward in energy. Similarly to previously discussed for the AA-stacked Lieb lattice systems in Sec.~\ref{BI_Lieb_lattice}, here the $N$-layered transition band structure consists of $N$ copies of monolayer TBs. Due to the non-null interlayer interaction, the $N$ TBs' copies will be shifted in energy with respect to each other, as can be seen in Figs.~\ref{espectro_transicao_mult}(a)-\ref{espectro_transicao_mult}(d). From the density of states shown in the right panels in Figs.~\ref{espectro_transicao_mult}(a)-\ref{espectro_transicao_mult}(d), one observes less pronounced peaks associated with the van-Hove singularities, and $N$ high significant peaks relative to the deformation of the $N$ flat bands around the zero energy level. Concerning the role of the next-nearest-neighbor interactions in the interlayer hopping parameters for the multilayer transition electronic properties, we present in Sec.~\textcolor{blue}{SIIIA} of the Supplemental Material \cite{SI} results comparing the band structures of bilayer [Fig.~\textcolor{blue}{S23}(b)] and trilayer [Fig.~\textcolor{blue}{S23}(e)] transition lattices and the density of states of the bilayer case in Fig.~\textcolor{blue}{S24}(b) when one considers solely perpendicular interlayer hoppings in the coupling Hamiltonian and when one also takes into account skewed interlayer hoppings. By comparing such spectra in Figs.~\textcolor{blue}{S23}(b) and \textcolor{blue}{S23}(e), one notes that the band structures exhibit only a slight energetic shift when one takes or not the skewed interlayer hoppings, as well as the van Hove singularity peaks in the density of states in Fig.~\textcolor{blue}{S24}(b). It demonstrates that a first-neighbor interlayer tight-binding model qualitatively and quantitatively captures accurately both the flat band physics and the general bands' format and curvatures without exhibiting drastic changes in the multilayer transition lattice band structures.

Analyzing the layer contribution in the probability density distributions for each energy band of the multilayer transition system in Fig.~\ref{espectro_transicao_mult}, which shows the band structures projected in the layers with the most significant probability densities (see colorbar scale), one observes a behavior for the electronic distribution with a sinusoidal pattern analogous to that of the AA-stacked multilayer Lieb lattice [Fig.~\ref{espectro_lib_mult}] as well as in a periodic superlattice of $N$ quantum wells, since the reflection symmetry between the layers is also preserved in the transition lattice. Thus, for the case of even $N$ layers, two layers always shall present the highest and the same contributions (blue), as can be seen in Figs.~\ref{espectro_transicao_mult}(a) and \ref{espectro_transicao_mult}(c) for $N=2$ and $N=4$, respectively. In contrast, for odd $N$ layers, there are cases in which one or more layers shall have nodal points in their electronic distributions. For example, for $N = 5$ in Fig.~\ref{espectro_transicao_mult}(c), the TBs in yellow are more located in the third layer and the TBs in purple are equally distributed in four layers, that are the first, second, fourth, and fifth layers, with the third layer acting as a nodal point for the electronic distribution. For the TBs in green in Fig.~\ref{espectro_transicao_mult}(d), it is observed that the second and fourth layers act as nodal points.

Similarly to the biased multilayer Lieb cases [Figs.~\ref{espectro_lib_mult}(e) and \ref{espectro_lib_mult}(f)], we next explore in Fig.~\ref{espectro_transicao_mult} the effect caused by a perpendicular electric field in the wave function distribution per layer when it is applied to the $N$-layered AA-stacked transition system. The presence of the electric field breaks the reflection symmetry of the system and consequently, it modifies the contribution per layer to the electronic states, which in turn, allows us to control the locations of the electronic states between the layers, as depicted in the bilayer and trilayer transition lattice band structures in Figs.~\ref{espectro_transicao_mult}(a) and \ref{espectro_transicao_mult}(e), and Figs.~\ref{espectro_transicao_mult}(b) and \ref{espectro_transicao_mult}(f), respectively, assuming a potential difference of (a, b) $\epsilon/|t| = 0$ and (e, f) $\epsilon/|t| = 1$. Similar to the multilayer Lieb case [Figs.~\ref{espectro_lib_mult}(e) and \ref{espectro_lib_mult}(f)], an imbalance in the electronic distribution and an energetic separation of the sets of TBs are observed in Figs.~\ref{espectro_transicao_mult}(e) and ~\ref{espectro_transicao_mult}(f) due to the breaking of the reflection symmetry caused by the non-null bias gate $\epsilon/|t| \neq 0$. The electrostatic bias causes the electronic states associated with the highest and lowest energy bands to become more localized in the outermost layers of the multilayer system. In addition, for the trilayer transition case [Fig.~\ref{espectro_transicao_mult}(f)], the eigenstates associated with the middle band tend to be located in the central layer ($2$nd layer), such as was observed for the $N=3$ Lieb case in Fig.~\ref{espectro_lib_mult}(f).

The role played by the bias on the eigenstates' $z$-distribution is also discussed in Sec.~\textcolor{blue}{IIA2} of the Supplemental Material \cite{SI}. As depicted in Fig.~\textcolor{blue}{S5}, the band structures of the bilayer transition lattice projected onto the first (top row of panels) and second layer (bottom row of panels) show that for $\epsilon/|t|=0$ both layers contribute equally per band, being this statement valid for all energy bands of the spectrum, \textit{i.e.} the probability density for all bands is equally distributed in both layers. A distinct situation is achieved when $\epsilon/|t| \neq 0$, which causes an electronic decompensation per band, that is, presenting a preferential accumulation of the probability density in one of the layers for a given band. Such behavior is analogous to that one obtained for the biased multilayer Lieb lattice shown in Fig.~\textcolor{blue}{S2}. Beyond that, a non-null $\epsilon$ value energetically shifts the two sets of TBs, as can also be verified in the density of states of the (a) bilayer and (b) trilayer AA-stacked transition systems in Fig.~\textcolor{blue}{S6}.

An alternative way of tuning the electronic properties and the distribution of wave functions between layers for each energy band is by varying the distances between adjacent layers in the multilayer system, which can be induced by hydrostatic pressure. \cite{da2024electronic} For this analysis, we show in Fig.~\textcolor{blue}{S7} of the Supplemental Material \cite{SI} (a) the energy bands as a function of the interlayer distance $d/a_0$ at the $\Gamma$--point for $N=2$ and $N=3$ cases, (b) the sublattice contributions for three specific energy states, and the AA-stacked bilayer bands structure for (c) $d/a_0<1$ and (d) $d/a_0>1$. As discussed in Sec.~\textcolor{blue}{SIIA2}, similar to the energy levels in two or three-quantum-well systems separated by a distance $d/a_0$, when they are far apart, their wave functions do not overlap and their energy levels are doubly and triply degenerate, respectively, whereas when they are closer such degeneracy is broken. It can be seen in Figs.~\textcolor{blue}{S7}(c) and \textcolor{blue}{S7}(d) for distances $d/a_0<1$ and $d/a_0>1$, in which the two sets of TBs for $N=2$ that are energetically far apart and tend to become doubly degenerate in the latter case, respectively. However, it is interesting to note in Fig.~\textcolor{blue}{S7}(a) that some states for the $N=3$ case remain unchanged (labeled by $E_{4,5,6}$) and are distributed in the outermost layers, \textit{i.e.} in the first and third layers.

In the interconvertibility context, the evolution of the AA-stacked bilayer transition band structures is depicted in Fig.~\textcolor{blue}{S8} of the Supplemental Material \cite{SI} from $\theta = 95^\circ$ to $\theta = 115^\circ$, as an example case. Similarly, as discussed in Sec.~\ref{sec.mono.transition} for monolayer transition lattice, for multilayer transition system in the absence of any external effects, one also observes any gap formation in the bands' evolution, presenting flat band deformation per set of TBs, which results into many titled cones in the energy spectra for $\theta \in (90^\circ ,120^\circ)$.

\subsection{Kagome lattice}\label{BI_Rede_Kagome}

\begin{figure*}[t]
    \centering
    \includegraphics[width=0.8\linewidth]{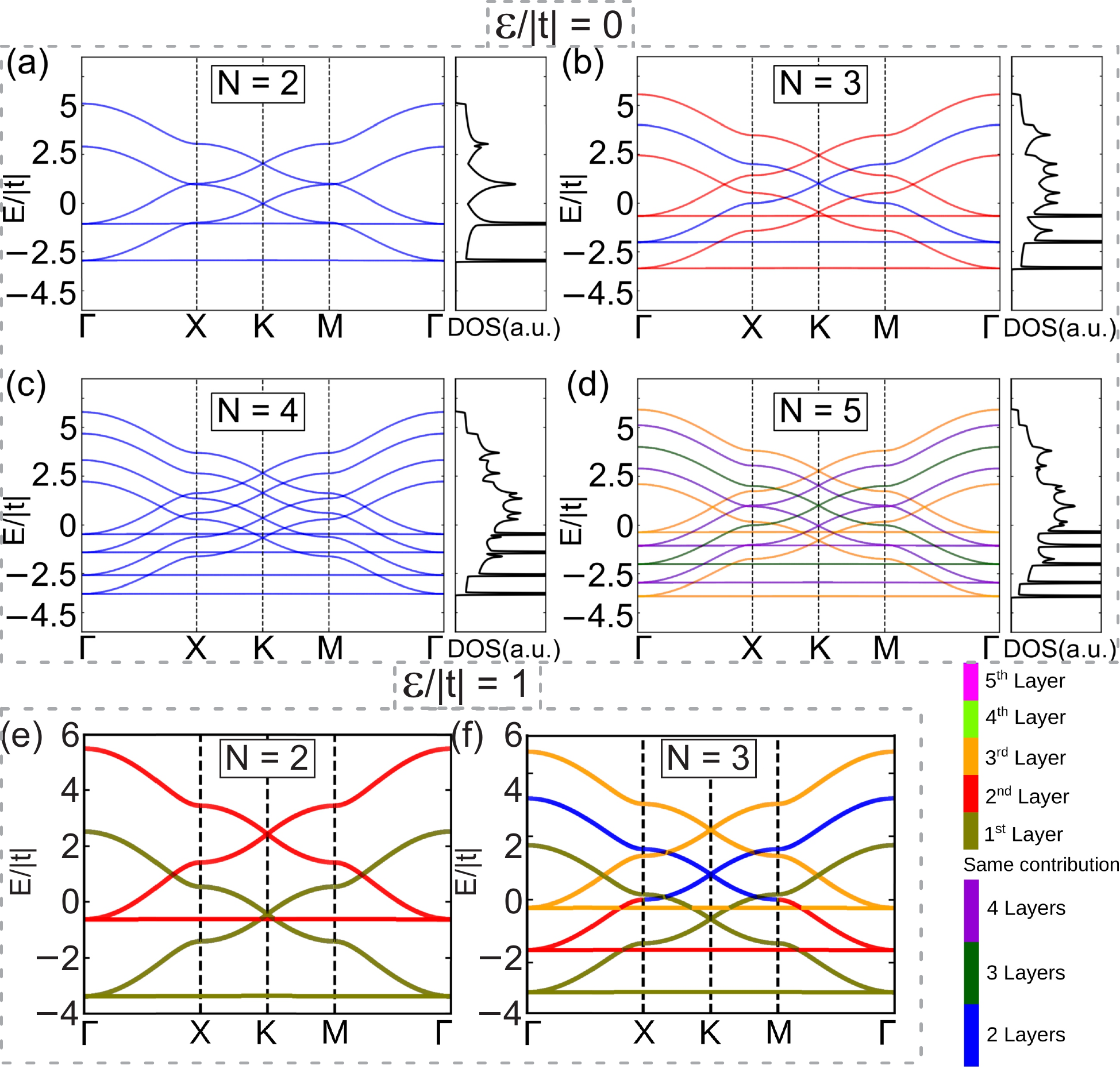}
    \caption{The same as in Fig.~\ref{espectro_lib_mult}, but now for AA-stacked multilayer Kagome lattice ($\theta = 2\pi/3$).}
    \label{espectro_kagome_mult}
\end{figure*}

The Kagome configuration corresponds to the last angle value of the evolutionary stage of the generic Lieb-Kagome lattice, when $\theta$ achieves the interatomic bond angle $\angle{ABC} \equiv 120^\circ$. After exploring the electronic aspects of the AA-stacked system formed by $N$ layers of Lieb (Sec.~\ref{BI_Lieb_lattice}) and transition (Sec.~\ref{BI_Rede_Transição}) lattices, we now focus on multilayer Kagome lattice systems with AA stacking \cite{PhysRevB.96.115426, crasto2019layertronic}. The multilayer Kagome band structures with $3N$ bands are obtained by diagonalizing the Hamiltonian \eqref{eqh13} with the coupling Hamiltonian elements for the $AA$ case given in Eq.~\eqref{eqh10}, taking for this current case $\theta = 2\pi /3$. Figure~\ref{espectro_kagome_mult} shows the energy spectra and the corresponding density of states for the Kagome lattice considering (a) $N=2$, (b) $N=3$, (c) $N=4$, and (d) $N=5$ layers. General electronic features discussed previously for Lieb and transition lattices with $N$ layers in Secs.~\ref{BI_Lieb_lattice} and \ref{BI_Rede_Transição}, respectively, are also clearly observed here for the Kagome case, \textit{e.g.}: (i) the $N$-layered Kagome band structure consists of $N$ copies of the monolayer Kagome bands (KB), being these copies shifted in energy due to the interaction between the layers, characterized by the interlayer hopping parameter, as can be observed in Figs.~\ref{espectro_kagome_mult}(a)-\ref{espectro_kagome_mult}(d) with two, three, four, and five sets of shifted KBs, respectively; (ii) the density of states for each case in Fig.~\ref{espectro_kagome_mult} presents smaller peaks associated with the van-Hove singularities and larger peaks associated with each flat band in the multilayer spectrum; (iii) when only perpendicular hoppings are taking into account in the interlayer coupling contributions between the adjacent layer, Figs.~\textcolor{blue}{S23}(c) and \textcolor{blue}{S23}(f) and Fig.~\textcolor{blue}{S24}(c) demonstrate that the multilayer Kagome band structures experience only a slight energetic shift, which is also reflected in an energetic displacement of the van Hove singularity peaks in the density of states.

Concerning the probability density distribution per layer and per band for the multilayer Kagome lattice, the colormap scale of the band structures in Fig.~\ref{espectro_kagome_mult} also carries information on layer-dependent wave function projection, accounting for the highest contribution at each $k$--point in the reciprocal space. Similar physical features as obtained for multilayer Lieb (Sec.~\ref{BI_Lieb_lattice}) and transition (Sec.~\ref{BI_Rede_Transição}) lattices with AA stacking demonstrated in Figs.~\ref{espectro_lib_mult} and \ref{espectro_transicao_mult}, here also due to reflection symmetry, one has (i) for the even number of layers that two layers always exhibit the highest contribution to the electronic states, as seen by the blue bands for all energy bands in Figs.~\ref{espectro_kagome_mult}(a) and \ref{espectro_kagome_mult}(c) for $N=2$ and $N=4$, respectively, whereas (ii) for $N$ odd, the probability density distribution among the layers follows a sinusoidal pattern, with certain layers acting as nodal points, as shown in Figs.~\ref{espectro_kagome_mult}(b) and \ref{espectro_kagome_mult}(d) for $N=3$ and $N=5$, respectively. By analogy, one observes that the electronic distribution, in the multilayer Kagome lattice, along the stacking direction follows a nodal pattern in a similar way as electronic states confined in an infinite quantum well with width $W = Nd$ or in a periodic potential like a Kronig Penney model formed by $N$ quantum wells \cite{crasto2019layertronic, ascroft, grosso2013solid}.

A physical analysis of the effects caused on the electronic properties of the multilayer AA-stacked Kagome lattice due to the application of an external electric field can be done by comparing the band structures in Figs.~\ref{espectro_kagome_mult}(a) and \ref{espectro_kagome_mult}(e) and Figs.~\ref{espectro_kagome_mult}(b) and \ref{espectro_kagome_mult}(f) for the bilayer and trilayer cases, respectively, taking a bias voltage of (a, b) $\epsilon/|t| = 0$ and (e, f) $\epsilon/|t| = 1$, as well as examining Figs.~\textcolor{blue}{S9} and \textcolor{blue}{S10} in Sec.~\textcolor{blue}{SIIA3} of the Supplemental Material \cite{SI} and the corresponding discussions for electronic layered-project bilayer Kagome band structures onto the first and second layers and the density of states for three different electric potential, respectively. As expected, breaking the reflection symmetry between the Kagome layers by applying a perpendicular electric field would cause changes in the relative contributions of each layer to the system's electronic states. As an example of how to incorporate the external electric field into the multilayer mode, one can refer to Eq.~\eqref{campo_1} for the total Hamiltonian of the bilayer case and now the current case taking $\theta = 2\pi/3$ into the $\hat{\hham}_{bi}$ term, resulting in an energy spectrum given by Eq.~\eqref{campo_2} in which $\varepsilon_{n}(\vec{k})$ shall correspond to the eigenvalues of the monolayer Kagome lattice. As observed for the AA-stacked multilayer Lieb [Figs.~\ref{espectro_lib_mult}(e) and \ref{espectro_lib_mult}(f)] and transition [Figs.~\ref{espectro_transicao_mult}(e) and \ref{espectro_transicao_mult}(f)] cases when an electrostatic potential difference is applied ($\epsilon/|t| \neq 0$), there is (i) an imbalance in the per-layer contributions, (ii) an increasing in the energy separation between the BKs \cite{crasto2019layertronic}, and (iii) a tendency for electronic states to be more localized in the outermost layers, as observed in Figs.~\ref{espectro_kagome_mult}(e) and \ref{espectro_kagome_mult}(f).

Regarding the effect of varying the distances between the adjacent layers, Fig.~\textcolor{blue}{S11} of the Supplemental Material \cite{SI} displays (a) the energetic dependence of the energy bands at the $\Gamma$-point as a function of the interlayer distance for $N=2$ and $N=3$, (c, d) the band structures for two different interlayer distances when it is $d/a_0<1$ and $d/a_0>1$, and (b) the sublattice contributions of the wave functions associated with the unchanged energies for the $N=3$ case. It is discussed in Sec.~\textcolor{blue}{SIIA3} the role of the interlayer distance in the band structures of $N=2$ and $N=3$-layered Kagome lattices in the analogy of a two- or three-level systems, namely, in the sense that reducing or increasing the interlayer distances is equivalent to increasing or decreasing the interlayer hopping energies between adjacent layers, leading to a higher or lesser energetic separation between the sets of KBs.

\section{AB-Stacking}\label{sec.IV}

The generic Lieb-Kagome lattice with the AB stacking can be obtained from the AA stacking by shifting one of the layers relative to the adjacent one along the in-plane $x$ direction by $\vec{a}_{1}/2$, as shown in Fig.~\ref{equema_bicamada}(b). In this type of stacking, non-dimer and dimer sites are formed. (Non)-Dimer sites are those that do (not) have direct perpendicular connections; that is, sites in the up/bottom layer that do (not) have a site immediately below at the bottom/up layer. Figures~\ref{equema_bicamada}(b) and \ref{equema_bicamada}(d) present the interlayer hopping parameters considered in our model between dimer sites and between the dimer and non-dimer sites. One observes for the AB-stacked multilayer Lieb-Kagome lattice that A and A$^\prime$ are non-dimer sites, whereas C (C$^\prime$) and B (B$^\prime$) are dimer sites, since the connection between the sites $B \leftrightarrow C^\prime$ ($C \leftrightarrow {B}^\prime$) occurs with the perpendicular hoppings being ${t}_{B{C}'}^{\perp}$ and ${t}_{C{B}'}^{\perp}$. Similar to the AA case, the multilayer Lieb-kagome lattice has $3N$ sites per unit cell, and consequently, $3N$ energy bands, obtained by diagonalizing the multilayer Hamiltonian given by Eq.~\eqref{eqh13}, considering the interaction Hamiltonian with the matrix elements for the AB stacking case given by Eq.~\eqref{eqh11}. As follows, Secs.~\ref{Lieb_AB}, \ref{Trasicao_AB}, and \ref{sec.Kagome.AB} address the electronic properties of Lieb $\left(\theta = \pi / 2\right)$, transition $\left(\pi/2 < \theta < 2\pi/3\right)$, and Kagome $\left(\theta = 2\pi/3\right)$ lattices, respectively, with the AB stacking, comparing the results, between the AA and AB stacking cases. More results and corresponding discussions for the electronic properties of multilayer generic Lieb-Kagome systems with AB stacking are presented in Secs.~\textcolor{blue}{SIIB} and \textcolor{blue}{SIIIB} of the Supplemental Material \cite{SI}.

\subsection{Lieb lattice}\label{Lieb_AB}

\begin{figure*}[t]
    \centering
    \includegraphics[width=0.8\linewidth]{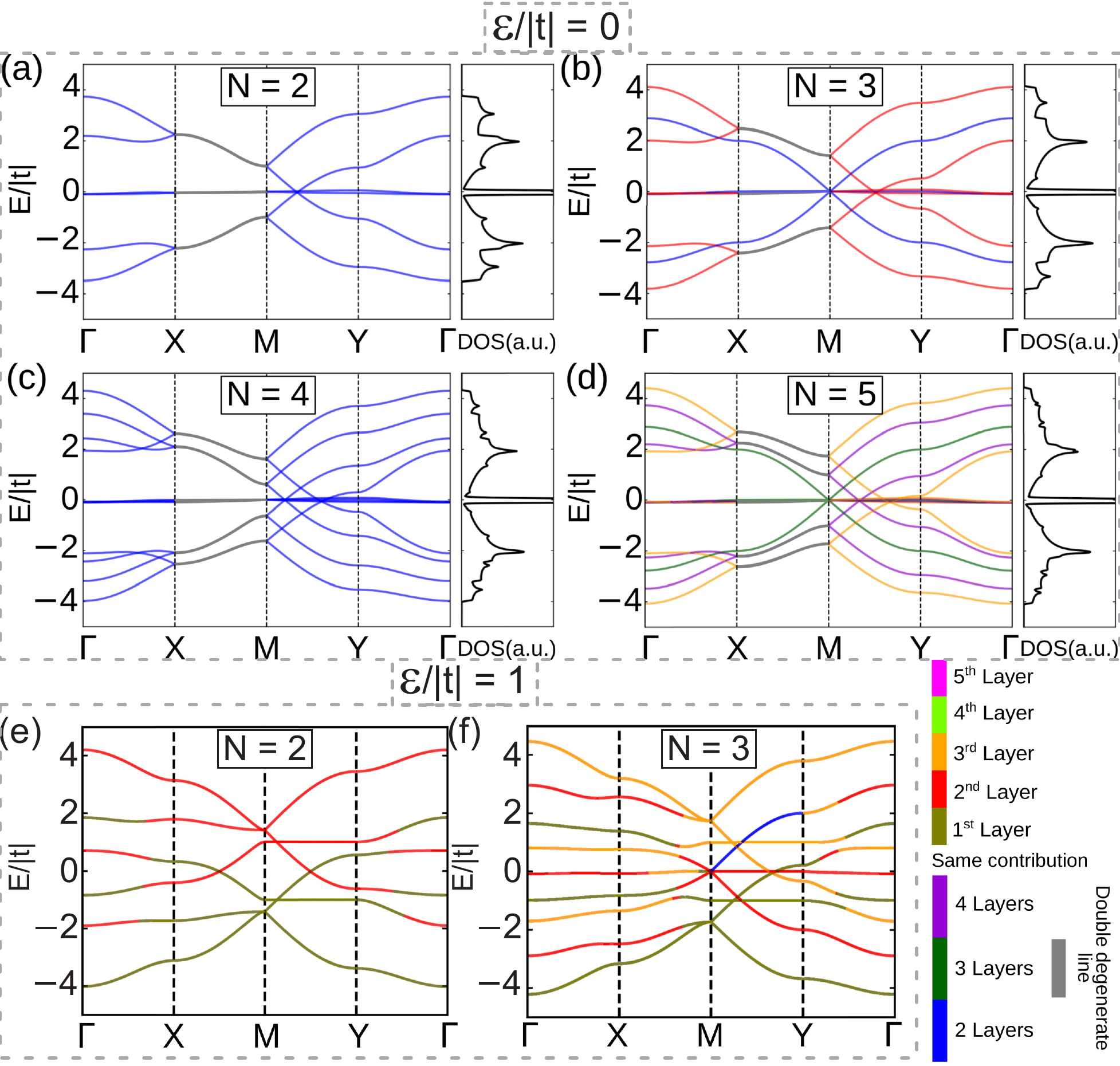}
    \caption{The same as in Fig.~\ref{espectro_lib_mult}, but now for multilayer Lieb lattice with AB stacking. Now, the color scheme also includes gray, which represents the doubly degenerate line.}
    \label{espectro_lib_mult_AB}
\end{figure*}

Figure~\ref{espectro_lib_mult_AB} presents the band structures and density of states of the multilayer Lieb lattice with AB stacking, considering (a) $N=2$, (b) $N=3$, (c) $N=4$, and (d) $N=5$ layers. Unlike the multilayer AA Lieb case discussed in Secs.~\ref{BI_Lieb_lattice} and \textcolor{blue}{SIIA1}, the AB-stacked $N$-layered Lieb band structures are not just $N$ identical copies of the monolayer Lieb case energetically displaced from each other due to the interlayer coupling, but rather present very distinct particularities. As can be observed for the bilayer case in Fig.~\ref{espectro_lib_mult_AB}(a), the $N=2$ band structure is double degenerate along the entire edge of the Brillouin zone in the path $\vec{X}-\vec{M}$, \textit{i.e.} three pairs of LBs labeled by the gray color, whereas along the other paths in the $k$ space, one observes six dispersive energy bands \cite{banerjee2021higher}. Moreover, it is interesting to note in Fig.~\ref{espectro_lib_mult_AB} for the $N$-layered Lieb lattice system with AB stacking, that the band structure along the paths $\vec{X}-\vec{M}$ and $\vec{M}-\vec{Y}$ in the Brillouin zone is asymmetric. This is seen by comparing Fig.~\ref{espectro_lib_mult} for the AA Lieb case with Fig.~\ref{espectro_lib_mult_AB} for the AB Lieb case. This asymmetry arises due to the breaking of the rotation symmetry $C_{4}$ in the AB configuration [Fig.~\ref{espectro_lib_mult_AB}], whereas in the AA stacking [Fig.~\ref{espectro_lib_mult}], $C_{4}$ operation is conserved, leading to symmetric band structures for the Lieb case along these paths. Delving deeper into the reason for such double degeneracy, one finds that it arises due to changing the symmetry space group to which it is assigned. The corresponding symmetry group of the AB stacking configuration is altered with respect to the AA-stacked case owing to a structural reason, since it is formed by a horizontal displacement of one of the layers from the adjacent ones in the in-plane direction. Thus, some of the symmetry operations must be combined with a non-primitive translation ($\vec{a}_{1}/2$) to keep the lattice invariant for the AB stacking case. Symmetry groups with this type of operation are known as non-symmorphic space groups, and one of the characteristics of these groups is the presence of additional degeneracies in the energy spectrum imposed by the symmetries \cite{dresselhaus2007group, zhao2016nonsymmorphic, young2015dirac, klemenz2020systematic, wu2022nonsymmorphic}.

Exploring electronic aspects of the band structures for an increasing number of layers, from Figs.~\ref{espectro_lib_mult_AB}(b)--\ref{espectro_lib_mult_AB}(d), one notices the emergence of new sets of LBs obeying different trends depending on whether $N$ is even or odd. For an odd number of layers, the resulting energy bands are composed by the combination of $(N-1)/2$ copies of the bilayer LBs and $1$ monolayer LB [see Figs.~\ref{espectro_lib_mult_AB}(b) and \ref{espectro_lib_mult_AB}(d) for $N=3$ and $N=5$, respectively]. This is the opposite of what happens in the AA-stacked Lieb case [Fig.~\ref{espectro_lib_mult}], in which, when adding an extra layer, a new set of LBs emerges, resulting in $N$ identical copies of LBs, regardless of parity of $N$. On the other hand, for an even number of layers, one observes a band structure composed of $N/2$ copies of the bilayer LBs, being therefore $3N/2$ bands doubly degenerate along the path $\vec{X}-\vec{M}$, as can be seen in Figs.~\ref{espectro_lib_mult_AB}(a) and \ref{espectro_lib_mult_AB}(c) for $N=2$ and $N=4$, respectively. In general, for $N$ even, the band structure is composed of bilayer-like bands, while for odd values of $N$, the structure is a composition of monolayer-like and bilayer-like bands. A similar effect has been observed in AB and ABA stacked graphene. \cite{sugawara2018selective, PhysRevB.82.035409, castroNeto, PhysRevB.78.205425, Castro_2010, McCann_2013, C5CP05013H, Min2012, PhysRevB.84.195453}


By evaluating the role played by the second-nearest-neighbor interlayer hoppings in the AB-stacked Lieb band structures for the bilayer [Fig.~\textcolor{blue}{S25}(a)] and trilayer [Fig.~\textcolor{blue}{S25}(d)] systems as well as the density of states for the $N=2$ case [Fig.~\textcolor{blue}{S26}(a)], one verifies by turning off these skewed interlayer hoppings that its absence reduces the dispersive character of LBs at the vicinity of $E/|t|=0$ and leads to flattening the LBs along the $\vec{M}-\vec{Y}$ path, whereas the density of states [Fig.~\textcolor{blue}{S26}(a)], when considered solely perpendicular interlayer hoppings, exhibts a energetic shift to high energies. Therefore, unlike the AA Lieb case discussed in Secs.~\ref{BI_Lieb_lattice} and \textcolor{blue}{SIIIA}, in which the general aspect of the Lieb band structure and density of states were not strongly affected by the second-nearest-neighbor interlayer hoppings, for the AB-stacked Lieb case they play a crucial role in the flatness of the quasi-flat bands, causing them to become flatter by turning off skewed interlayer hoppings. In fact, such connection between the quasi-flat bands and the strength of the second-nearest-neighbor interlayer hoppings could be achieved by taking a larger $n$ value in the hopping function given by Eq.~\eqref{hopping1}, which would lead to a faster position-decaying energetic tendency and thus practically turning off the contributions of second interlayer neighbors, similarly as was discussed in Refs.~[\onlinecite{PhysRevB.108.125433, wellissonTPT}] for intralayer hoppings in the Lieb-Kagome lattice and in Sec.~\textcolor{blue}{SI} of the Supplemental Material \cite{SI}. The flatness of the energy bands shall impact the transport and optoelectronic properties of AB-stacked multilayer Lieb systems. Additional discussions concerning the interlayer hopping contributions for the AB-stacked multilayer Lieb system are presented in Sec.~\textcolor{blue}{SIIIB} of the Supplemental Material \cite{SI}.

In the context of the layer-dependence on the location of the electronic states, the multilayer Lieb band structures shown in Fig.~\ref{espectro_lib_mult_AB} also inform about the electronic contributions projected per layer, labeling the states according to the highest contribution per energy band in a similar way as the same color scheme used for the AA stacking case in Fig.~\ref{espectro_lib_mult}; excepted by the double-degenerate bands along the $\vec{X}-\vec{M}$ path that is labeled now in gray color and whose probability density's contributions per layer shall be discussed as follows here. For $N$ even, the bilayer and tetralayer AB-stacked Lieb band structures presented in Figs.~\ref{espectro_lib_mult_AB}(a) and \ref{espectro_lib_mult_AB}(c), respectively, show that almost the whole energy bands, except along the $\vec{X}-\vec{M}$ path, have the two layers contributing equally to the electronic states (blue curves). From the color scheme in Figs.~\ref{espectro_lib_mult_AB}(b) and \ref{espectro_lib_mult_AB}(d) for $N=3$ and $N=5$, respectively, one notices for odd $N$ layers that the electronic distribution of states along the $z$ direction follows the sinusoidal behavior as Lieb case with AA stacking in Sec.~\ref{BI_Lieb_lattice}. For instance, for $N=3$, one has a set of six bands with a higher contribution of the wave function distribution per layer in the second layer (red curves), and another set of three bands in which two layers contribute equally (blue curves). By examining the layer-dependence of the electronic distribution for the double-degenerate bands along the $\vec{X}-\vec{M}$ path labeled in gray in Fig.~\ref{espectro_lib_mult_AB}, a distinct behavior was observed. In Fig.~\textcolor{blue}{S12} of the Supplemental Material \cite{SI}, the double-degenerate bands along the $\vec{X}-\vec{M}$ path are presented for the [Fig.~\textcolor{blue}{S12}(a)] bilayer and [Fig.~\textcolor{blue}{S12}(e)] trilayer Lieb lattice cases with the AB stacking, along with the wave functions for $N=2$ in Figs.~\textcolor{blue}{S12}(b)-\textcolor{blue}{S12}(d) and $N=3$ in Figs.~\textcolor{blue}{S12}(f)-\textcolor{blue}{S12}(h) for three different $k_{y}$ values. It demonstrates that such double-degenerate states along the $\vec{X}-\vec{M}$ path exhibit strong layer polarization near $\vec{X}$, which gradually decreases towards the $\vec{M}$ point, indicating a redistribution of the electronic density among the layers without the formation of nodal points. For instance, compare the layer-projected wave function distribution in Figs.~\textcolor{blue}{S12}(b)-\textcolor{blue}{S12}(f) for $N=2$ and $N=3$ at $k_{1y}$-point, being them preferentially located at one layer, whereas one approaches to the $\vec{M}$, such at $k_{3y}$-point with wave function shown in Figs.~\textcolor{blue}{S12}(d)-\textcolor{blue}{S12}(h), one has a delocalization of the wave functions.

Figures~\ref{espectro_lib_mult_AB}(a) and \ref{espectro_lib_mult_AB}(e) and Figs.~\ref{espectro_lib_mult_AB}(b) and \ref{espectro_lib_mult_AB}(f) show the band structures of the (a, c) bilayer and (b, d) trilayer Lieb lattice with the AB stacking, considering a potential difference of (a, b) $\epsilon/|t| = 0$ and (e, f) $\epsilon/|t| = 1$. When a perpendicular electric field is applied to the multilayer AB-stacked Lieb system, significant changes are observed in the band structure, causing a redistribution of the electronic states per layer and lifting degenerate states. For instance, for the bilayer case shown in Figs.~\ref{espectro_lib_mult_AB}(a) and \ref{espectro_lib_mult_AB}(e), the degeneracy is lifted along the path $\vec{X}-\vec{M}$, along with a redistribution of the states per layer, as seen by the changes in the colors of the bands. Due to the potential difference, some symmetry operations of the nonsymmorphic space group are broken, and consequently, the degeneracy of the electronic bands protected by these symmetries is lifted \cite{young2015dirac, yang2019topological}.

Unlike the AA stacking case (Sec.~\ref{BI_Lieb_lattice}), the application of an electrostatic potential difference of intensity $\epsilon/|t| =1$ for the AB-stacked case does not cause the bands to be fully polarized per layer. Due to the presence of dimer and non-dimer sites, the bands are partially polarized in different layers. It can be seen from Fig.~\ref{espectro_lib_mult_AB}(e) that the band crossings in the path $\vec{M}-\vec{Y}$ lead to the formation of tilted cones close to the $\vec{M}$-point and a cone formed in the middle way between $\vec{M}$ and $\vec{Y}$ points, composed by states located in different layers with the state with positive (negative) group velocity located at the first (second) layer. Such behavior is not specific for $N=2$ [Fig.~\ref{espectro_lib_mult_AB}(e)], but also occurs for the trilayer case as viewed by comparing Figs.~\ref{espectro_lib_mult_AB}(b) and \ref{espectro_lib_mult_AB}(f). In addition, the triply degenerate point observed in the trilayer case at $\vec{M}$ point is not affected by the application of the electric field. Additional results in the presence of perpendicular electric field are shown in Sec.~\textcolor{blue}{SIIB1} for [Fig.~\textcolor{blue}{S13}] the bilayer Lieb band structure projected onto the first and second layers and [Fig.~\textcolor{blue}{S14}] the density of states for both the bilayer and trilayer systems for three different bias gate values. These results demonstrate that the electrostatic potential difference strongly modifies the density of states, splitting the peaks in both $N=2$ and $N=3$ cases. However, a sharp peak is still observed even at high electric field magnitudes around $E/|t| \approx 0$ for the trilayer Lieb case as a result of the flatness of the bands around that energy, as can be seen in Fig.~\ref{espectro_lib_mult_AB}(f). A notable aspect in the density of states of the AB-stacked Lieb case that differentiates from the AA-stacked Lieb one when the electric field is applied is that rather the density of states to exhibit multiple peaks and the applied electric field only induces a shift in these peaks as happens in AA stacking; conversely, in AB-stacked Lieb case, the density of states is dominated by a prominent peak at $E\approx 0$, which broadens significantly upon application of an electric field. This distinct response is anticipated to have significant implications for various electronic properties of the multilayer AB-stacked Lieb lattice. Moreover, by comparing the AA-stacked [Fig.~\textcolor{blue}{S2}] and AB-stacked [Fig.~\textcolor{blue}{S13}] bilayer Lieb layer-projected band structures, one can realize that bilayer LBs are not fully layered-polarized under the effect of bias gates throughout the whole Brillouin zone, but instead, some bands show per-layer hybridization, \textit{i.e.} mixtures of the wave function contribution being localized in both layers, especially in the energetic regions close to the anti-crossings, as can be verified the colormap of the bands at $\vec{Y}$ and $\vec{X}$ points in Figs.~\textcolor{blue}{S13}(c) and \textcolor{blue}{S13}(f). Regarding the system's electronic properties, when one varies the interlayer distance, Fig.~\textcolor{blue}{S15} displays the band structure for different distance values and the corresponding wave functions. For large distances, the energy bands tend to converge to the monolayer LBs, whereas for small distances, low-energy bands emerge, primarily governed by non-dimer sites. Furthermore, reducing the interlayer distance can increase the number of quasi-flat bands, suggesting that vertical strains, such as hydrostatic pressure, may serve as a mechanism to induce the emergence of more flat bands in the electronic spectrum of multilayer AB-stacked Lieb systems.

\subsection{Transition lattice}\label{Trasicao_AB}

\begin{figure*}[t]
    \centering
    \includegraphics[width=0.8\linewidth]{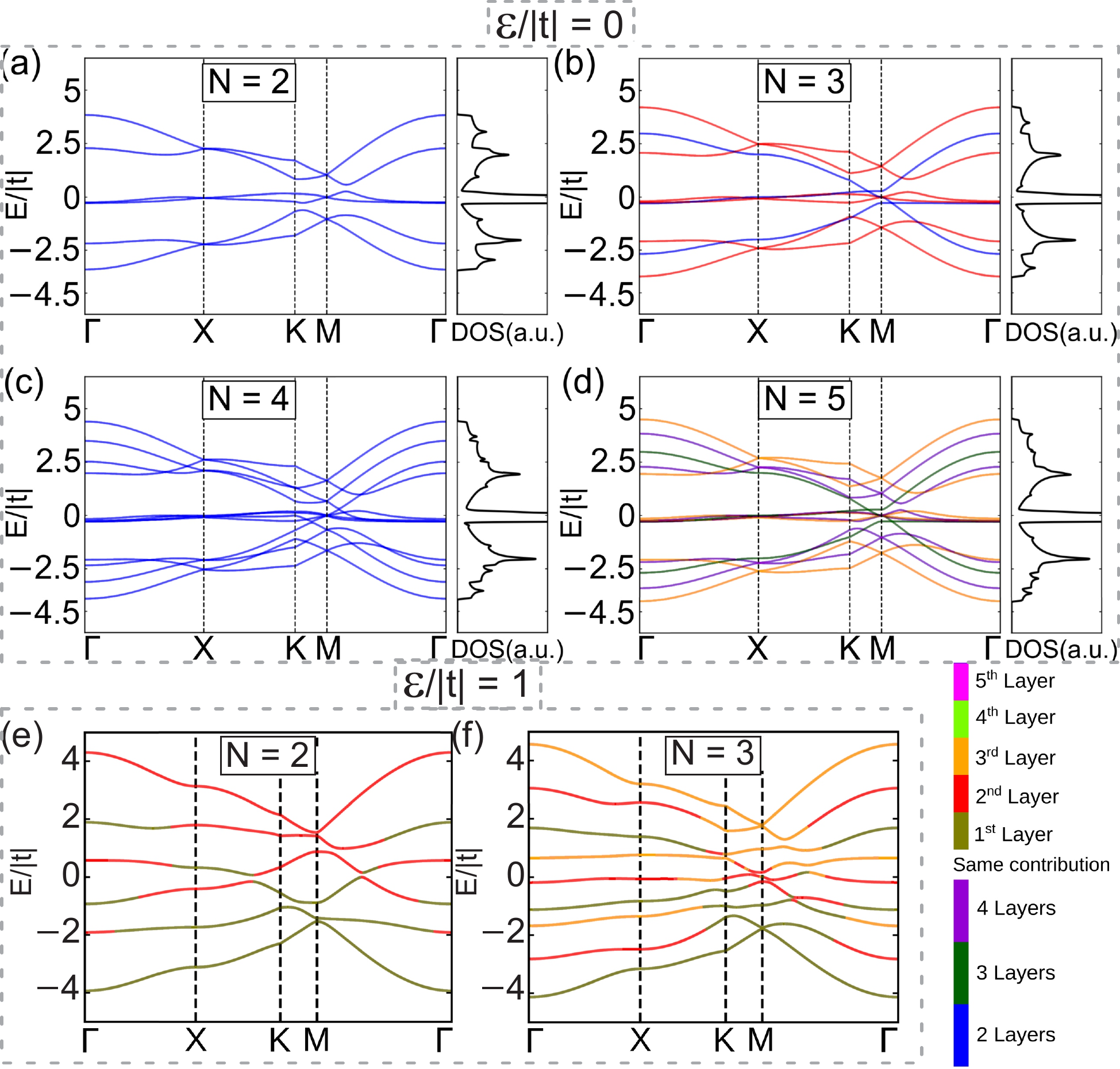}
    \caption{The same as in Fig.~\ref{espectro_lib_mult}, but now for multilayer transition lattice with AB stacking assuming $\theta = 7\pi/12$.}
    \label{espectro_transicao_mult_AB}
\end{figure*}

The study of the electronic properties of the multilayer transition lattice with the AB stacking is carried out in this section through the diagonalization of the multilayer Hamiltonian [Eq.~\eqref{eqh13}] with the interlayer couplings given by Eqs.~\eqref{eqh11}(a)-\eqref{eqh11}(d) that account properly for the hopping energies between the two dimer sites and the two non-dimer sites formed because of the translation of the upper layer into half the value of the lattice parameter along the horizontal direction in the AB-stacked configuration [Figs.~\ref{equema_bicamada}(b) and \ref{equema_bicamada}(d)]. Due to this topology in the unit cell, one expects that the electronic properties of multilayer AB-stacked transition lattices exhibit a more distinct behavior compared to the monolayer and the AA-stacked cases of the transition lattice (Sec.~\ref{BI_Rede_Transição}). This can be verified by analyzing Fig.~\ref{espectro_transicao_mult_AB}, that shows the multilayer transition lattice band structures considering $\theta = 7\pi/12$, as an example case of transition lattice, and taking (a) $N=2$, (b) $N=3$, (c) $N=4$, and (d) $N=5$ layers.

The band structure of the bilayer AB-stacked transition lattice [Fig.~\ref{espectro_transicao_mult_AB}(a)] exhibits six dispersive bands along the entire path shown here in the $k$-space. Banerjee and Saxena~\cite{banerjee2021higher} demonstrated that the doubly-degenerate energetic lines for the Lieb lattice in the AB and ABC stacking are lost when subjected to some deformation. However, it is worth emphasizing the difference between their tight-binding model and ours concerning the interlayer hoppings. While in the present model, the interactions are position-dependent [Eq.~\eqref{hopping1}], in contrast, Banerjee \& Saxena's model \cite{banerjee2021higher} assumes fixed values for some interlayer hoppings, whereas for other interlayer connections, an interpolation between the hopping parameters of the limiting lattices is considered. Within this theoretical approach, they observed that the states associated with the double-degenerate energetic lines have their degeneracy lifted for $\theta \neq \pi/2$ together with an opening of a local gap. In this perspective, the model proposed here is more general, and allows us to obtain results similar to those of the literature\cite{banerjee2021higher} without any arbitrariness in the choice of the system parameters, but instead obeying an intuitive position-dependent relation for the interatomic bonds.

To better visualize such evolution from non-dispersive to dispersive LBs in the interconvertibility process, we show in Fig.~\textcolor{blue}{S16} in Sec.~\textcolor{blue}{SIIB2} of the Supplemental Materials \cite{SI} the bilayer AB-stacked transition lattice band structure evolution through the Lieb to Kagome morphology for different angles $\theta$. Moreover, it is interesting to observe, as indicated by black circles in Fig.~\textcolor{blue}{S16}, that the energetic degeneracy at the $\vec{X}$ and $\vec{M}$ points are preserved due to the maintenance of some symmetry operations that are common among the lattice evolution. Such band evolution also leads to the AB-stacked band structure around $\vec{K}$ point to be symmetric, as can be seen by comparing the TBs' dispersion for the different $\theta$ values in Fig.~\textcolor{blue}{S16} along the $\vec{X}-\vec{K}$ and $\vec{K}-\vec{M}$ paths.

Note from Figs.~\ref{espectro_transicao_mult_AB}(a) and \ref{espectro_transicao_mult_AB}(c), that for even $N$, the AB-stacked TBs contain $N/2$ copies of the energy-shifted bilayer TBs, emerging new crossing points between the set of TBs. For the case of odd $N$, the multilayer AB-stacked transition band structure is composed of a combination of $(N-1)/2$ copies of the energy-shifted bilayer TBs and a copy of the monolayer transition energy spectrum, as can be seen in Figs.~\ref{espectro_transicao_mult_AB}(b) and \ref{espectro_transicao_mult_AB}(d). When evaluating the role of second-neighbor interlayer hoppings on the AB-stacked transition band structures [Fig.~\textcolor{blue}{S25}] and the density of states [Fig.~\textcolor{blue}{S26}] for the bilayer [Fig.~\textcolor{blue}{S25}(b)] and trilayer [Fig.~\textcolor{blue}{S25}(e)] cases, only a slight shift in energy is observed, meaning that a simplified first-nearest-neighbor interlayer tight-binding model is be able to capture the main features of the spectrum during the interconvertibility process for multilyaer AB-stacked transition lattice. More details are presented in Sec.~\textcolor{blue}{SIIIB} of the Supplemental Materials \cite{SI}.

\begin{figure*}[t]
    \centering
    \includegraphics[width=0.8\linewidth]{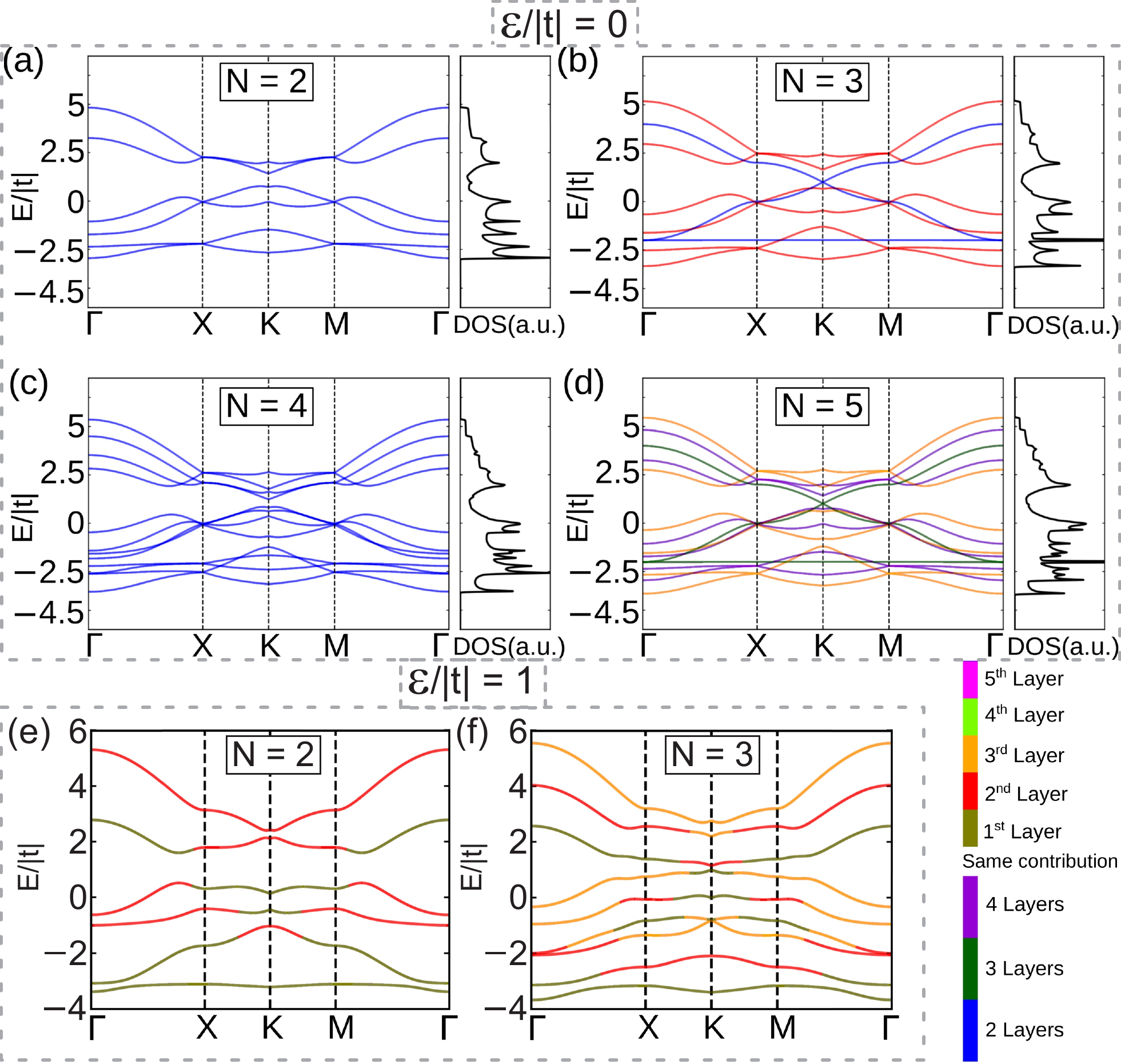}
    \caption{The same as in Fig.~\ref{espectro_lib_mult}, but now for multilayer Kagome lattice ($\theta = 2\pi/3$) with AB stacking.}
    \label{espectro_kagome_mult_AB}
\end{figure*}

The color scheme in Fig.~\ref{espectro_transicao_mult_AB} denotes the wave function distribution per layer, indexing the layer in which the electronic state has a high localization amplitude. One observes that the distribution of the states per layer follows a sinusoidal behavior along the $z$ direction, similar to all previous cases. For the degenerate states present in the spectrum, the distribution maintains the same pattern observed for the doubly degenerate line in the AB-stacked Lieb lattice shown in Fig.~\ref{espectro_lib_mult_AB}, with each state distributed in different layers. The effect of the applied electric field on the band structure and wave function distribution of the multilayer AB-stacked transition system can be verified by comparing Figs.~\ref{espectro_transicao_mult_AB}(a) and \ref{espectro_transicao_mult_AB}(e), and Figs.~\ref{espectro_transicao_mult_AB}(b) and \ref{espectro_transicao_mult_AB}(f), where we present the band structures for the (a, e) bilayer and (b, f) trilayer transition lattices with AB stacking, respectively, for two values of potential difference: (a, b) $\epsilon/|t| = 0$ and (e, f) $\epsilon/|t| = 1$. Concerning such influence on the distribution of the states and their locations in different layers, one can observe for the bilayer case [Figs.~\ref{espectro_transicao_mult_AB}(a) and \ref{espectro_transicao_mult_AB}(e)] the separation of the degenerate states that it is due to the breaking of the non-symmorphic symmetry operations, \textit{i.e.}, a degeneracy breaking in addition to the redistribution of the electronic states in the layers. As discussed in Sec.~\ref{Lieb_AB} and observed in Fig.~\ref{espectro_lib_mult_AB} for the AB-stacked Lieb lattice, some energy bands are partially distributed in different layers as a consequence of the reduction in the number of interlayer hoppings compared to the transition lattice case with the AA stacking. The total polarization of the states per layer occurs only for potential difference values greater than the perpendicular hopping. A similar behavior is observed for the trilayer case [Figs.~\ref{espectro_transicao_mult_AB}(b) and \ref{espectro_transicao_mult_AB}(f)], with the bands partially distributed between the layers. By comparing the effects of the applied bias gate on the band structures for $N=2$ [(Fig.~\ref{espectro_lib_mult_AB}(e)) Fig.~\ref{espectro_transicao_mult_AB}(e)] and $N=3$ [(Fig.~\ref{espectro_lib_mult_AB}(f)) Fig.~\ref{espectro_transicao_mult_AB}(f)] for (Lieb) transition lattices with AB stacking, it is seen that the bands around $E=0$ become less dispersive for $N=3$ and AB-stacked transition cases. To check the wave functions' projection on the TBs, we show in Fig.~\textcolor{blue}{S17}, as an example case, the AB-stacked bilayer band structure of the transition lattice under three different bias voltages, in which the colormap scale denotes its projection onto the first and second layers. Due to the emergence of anti-crossings in the band structure along the shown $k$-space paths, one realizes in Fig.~\textcolor{blue}{S17} wave function per-layer hybridizations, demonstrating, in turn, a more complex symmetry of the AB-stacked transition system and its uncoupled nature in two simple monolayer transition lattices in comparison to the AA-stacked bilayer transition lattice, as demonstrated in Fig.~\textcolor{blue}{S5} that can be layer-polarized by the application of a perpendicular electric field. In summary, it means that one can not easily interpret such a system as solely two shifted sets of TBs to form the energy spectrum of the AB-stacked bilayer transition lattice. Such a complex character of the AB-stacked transition band structure will also reflect in a complex density of states, as can be seen in Figs.~\textcolor{blue}{S18}(a) and \textcolor{blue}{S18}(b) for bilayer and trilayer AB-stacked transition systems, respectively. One realizes, by applying a potential difference, that the split of the energy bands increases, and consequently, it leads the density of states to have more peaks with smaller amplitudes, which are then shifted, leading to the reduced number of states near the zero energy level. For the trilayer case, Fig.~\textcolor{blue}{S18}(b) also shows a sharp peak near zero energy. Another way to modify the AB-stacked transition electronic band structure is by varying the interlayer distances between the adjacent layers. With respect to that, we show in Fig.~\textcolor{blue}{S19} the energy levels at $\vec{\Gamma}$ point as a function of the interlayer distance $d/a_0$ for $N=2$ and $N=3$ [Fig.~\textcolor{blue}{S19}(a)], the corresponding wave functions for the unalterated states when $d/a_0$ changes and for those that do not diverge at $d/a_0<1$ for $N=2$ [Fig.~\textcolor{blue}{S19}(b)] and $N=3$ [Fig.~\textcolor{blue}{S19}(c)], and the band structures for $d/a_0<1$ [Fig.~\textcolor{blue}{S19}(d)] and $d/a_0>1$ [Fig.~\textcolor{blue}{S19}(e)]. For large interlayer distances ($d/a_0>1$), the energy bands tend to converge to the monolayer transition band structure with the states becoming $N$-fold degenerate, whereas for small distances ($d/a_0<1$), the low-energy bands break their degeneracy, with their wave functions revealing to be contributions primarily governed by non-dimer sites. More discussions are presented in Sec.~\textcolor{blue}{SIIB2} of the Supplemental Materials \cite{SI}.

\subsection{Kagome lattice}\label{sec.Kagome.AB}

In this section, we discuss the electronic properties of the AB-stacked multilayer Kagome lattice, analogously to the previous Secs.~\ref{Lieb_AB} and \ref{Trasicao_AB} in which we explored the AB-stacked Lieb and transition cases. The atomic sites here are organized into dimer and non-dimer sites, with the unit cell containing $3N$ sites, where $N$ is the number of layers, resulting in a band structure consisting of $3N$ bands. Such results considering (a) $N=2$, (b) $N=3$, (c) $N=4$, and (d) $N=5$ layers of Kagome lattice in the AB stacking are depicted in Fig.~\ref{espectro_kagome_mult_AB}. Similarly to the observed electronic feature for the AB-stacked transition lattice in Fig.~\ref{espectro_transicao_mult_AB}(a) in Sec.~\ref{Trasicao_AB} for the bilayer case, Fig.~\ref{espectro_kagome_mult_AB}(a) for $N=2$ Kagome lattice exhibits six dispersive bands being doubly-degenerate in pairs at the $\vec{X}$ and $\vec{M}$ points. This is a consequence of non-symmorphic symmetry of the system in the AB arrangement \cite{PhysRevB.100.155421,crasto2019layertronic}. It should be noted that the gap presented in the bilayer band structure does not correspond to a global gap, as can be observed in the density of states. For the trilayer case, Fig.~\ref{espectro_kagome_mult_AB}(b) shows that the AB-stacked Kagome band structure is composed of a combination of the bilayer KBs, presenting therefore doubly-degenerate states in pairs at the $\vec{X}$ and $\vec{M}$ points, and the monolayer KBs. For even $N$, as exemplified for Fig.~\ref{espectro_kagome_mult_AB}(c) for $N = 4$, the band structure features shifted copies in energy of the bilayer KBs. In contrast, for odd $N$ layers, the AB-stacked Kagome band structure consists of a combination of $(N-1)/2$ copies of the bilayer KBs shifted in energy and a copy of the monolayer KBs, as shown in Fig.~\ref{espectro_kagome_mult_AB}(d) for $N = 5$. Unlike the previously analyzed AB-stacked Lieb (Fig.~\ref{espectro_lib_mult_AB}) and transition (Fig.~\ref{espectro_transicao_mult_AB}) lattice cases, in which the $\vec{K}$ point was not diametrically in the middle of the $\vec{X}-\vec{K}-\vec{M}$ path and the energy bands were not symmetrical with respect to $\vec{K}$ point in the first Brillouin zone, now for AB-stacked Kagome lattice one has a symmetric band structure in relation to the high symmetry point $\vec{K}$, \textit{i.e.} the AB-stacked Kagome band structure obeys a energetic-mirror-symmetry with respect to the $\vec{K}$ point, being equivalent in the $\vec{\Gamma}-\vec{X}-\vec{K}$ and $\vec{K} - \vec{M} - \vec{\Gamma}$ paths.

In the context of the wave function distribution per layer and per band, \textit{i.e.} the eigenstates' localization, in Fig.~\ref{espectro_kagome_mult_AB}, each band was labeled according to a color scheme to indicate in which layer each state presents a high contribution of its probability density. From that, it can be seen that the electronic distribution per layer follows a sinusoidal pattern for $N$ odd, analogous to infinite well states. The degenerate states maintain the distribution observed in the doubly degenerate energetic line of the AB-stacked Lieb lattice in Fig.~\ref{espectro_lib_mult_AB}, with each state localized in different layers. When analyzing the impact of second-neighbor interlayer hoppings in the AB-stacked bilayer [Fig.~\textcolor{blue}{S25}(c)] and trilayer [Fig.~\textcolor{blue}{S25}(f)] systems, one notices only a minor energy shift in the band structures, and a similar energetic shift effect is also seen in the density of states, as shown in Fig.~\textcolor{blue}{S26}(b) of the Supplemental Materials \cite{SI}. More discussions on that are presented in Sec.~\textcolor{blue}{SIIIB} of the Supplemental Materials \cite{SI}.

Regarding the effect of the perpendicular electric field on the
band structure of the AB-stacked Kagome lattice, we show, as two example cases, in Figs.~\ref{espectro_kagome_mult_AB}(a) and \ref{espectro_kagome_mult_AB}(e) and Figs.~\ref{espectro_kagome_mult_AB}(b) and \ref{espectro_kagome_mult_AB}(f), respectively, results for the bilayer and trilayer systems for two different values of potential difference: (a, b) $\epsilon/|t|=0$ and (e, f) $\epsilon/|t|=1$. For the bilayer case, Figs.~\ref{espectro_kagome_mult_AB}(a) and \ref{espectro_kagome_mult_AB}(e) show that when a potential difference is applied, the two-fold degeneracies at the $\vec{X}$ and $\vec{M}$ points are lifted due to the breaking of the non-symmorphic symmetries. As previously observed for the other lattices, a redistribution of electronic states per layer is also observed here for the multilayer AB-stacked Kagome case, with bands partially distributed in different layers. Such general behavior of the degeneracy breaking, band splitting, and band flatness induced by a non-null bias voltage is also observed for the trilayer case as shown by comparing the band structure in Fig.~\ref{espectro_kagome_mult_AB}(b) for $\epsilon/|t|=0$ and Fig.~\ref{espectro_kagome_mult_AB}(f) for $\epsilon/|t|=1$. An analysis of the AB-stacked bilayer Kagome band structure under the influence of three different electric field amplitudes is performed in Sec.~\textcolor{blue}{SIIB3} of the Supplemental Materials \cite{SI}. Figure~\textcolor{blue}{S20} in the Supplemental Materials \cite{SI} shows such AB-stacked bilayer band structures with the colormap scale associated with the layer-projection onto the first and second layers for three different potential differences. It demonstrated that the electronic bands with lower and higher energies are more likely to be electrostatically per-layer polarized, as can be verified by the low and high color intensities for the first and the sixth energy bands, respectively, when they are projected onto the second [Figs.~\textcolor{blue}{S20}(d) - \textcolor{blue}{S20}(f)] and first [Figs.~\textcolor{blue}{S20}(a) - \textcolor{blue}{S20}(c)] layers. The apparent anti-crossings in the chosen $k$-space path ($\vec{\Gamma} - \vec{X} - \vec{K} - \vec{M} - \vec{\Gamma}$) in the first Brillouin zone present greater hybridizations per layer per band, showing a mixture of the wave function contributions in both layers of the AB-stacked bilayer system. It is worth emphasizing that no total gap is opened in the system, and such anti-crossings are solely local and apparent gaps. The effect of the electric field on the AB-stacked bilayer and trilayer Kagome systems can be verified in the density of states depicted in Fig.~\textcolor{blue}{S21} of the Supplemental Materials \cite{SI}. As a consequence of the applied potential difference to the AB-stacked Kagome system, which leads to the innermost energy bands to become less dispersive, this results in the emergence of new peaks in the density of states with smaller amplitudes when compared to the $\epsilon/|t|=0$ case, but it does not open a gap between the bands, despite the symmetry breaking caused by the electric field. Finally, Fig.~\textcolor{blue}{S22} displays the band structure and the corresponding wave functions for varying the distances between the Kagome layers. As discussed previously for the analyzed multilayer systems here, for large distances ($d/a_0>1$), the energy bands converge to the monolayer KBs, whereas for smaller distances ($d/a_0<1$), the degeneracy of the low-energy bands is lifted, mainly driven by non-dimer sites. More discussions on that are presented in Sec.~\textcolor{blue}{SIIIB} of the Supplemental Materials \cite{SI}.

\section{Conclusions}\label{sec.conclusions}

In summary, we systematically studied the electronic properties of the multilayer Lieb-Kagome lattice system with the AA and AB stackings. Based on a generic Hamiltonian that describes the interconvertibility process between the monolayer Lieb and Kagome lattices proposed in Ref.~[\onlinecite{tony2019}], we extended such a tight-binding model for the multilayer Lieb-Kagome case, adopting a block tridiagonal $N \times N $ Hamiltonian. In this model, the main diagonal elements correspond to the monolayer Lieb-Kagome Hamiltonian, which depends on the angle $\theta$ associated with the Lieb-Kagome lattice morphology, while the off-diagonal elements represent the interlayer interactions, whose form depends on the considered stacking.

For the AA stacking, the reflection symmetry between layers results in $N$ energy-shifted copies of the monolayer Lieb-Kagome band structure, a behavior observed for all three studied lattices. In contrast, due to the unit cell topology, which includes dimer and non-dimer sites, for AB stacking, we identified the formation of degenerate states at the $\vec{X}$ and $\vec{M}$ points. This degeneracy arises from the transition of the system's space group classification, changing from symmorphic to non-symmorphic.

Analyzing the spatial distribution of the electronic states, we observed a sinusoidal behavior characteristic of states confined in an infinite potential well of width $L = (N+1)d$, where $d$ is the interlayer distance and $N$ is the number of layers. However, in the AB stacking, the degenerate states imposed by nonsymmorphic symmetries exhibit layer polarization. The application of an electric field breaks the reflection symmetry between the layers, leading to an imbalance in the electronic state distribution. In AA stacking, the electric field merely shifts the electronic bands without altering their energetic shape, whereas in AB stacking, it redistributes the states across the layers, lifts the degeneracy of nonsymmorphic states, and increases band dispersion.

From the theoretical analysis developed in this study, we can conclude that the electronic properties of the Lieb, transition, and Kagome multilayer lattices are highly sensitive to the stacking type and the number of layers, directly influencing the wave function distribution and the system's response to external perturbations such as an applied electric field. Moreover, the extended tight-binding model for the multilayer Lieb-Kagome lattice based on a generic one-parameter ($\pi/2 \leq \theta \leq 2\pi /3$) Hamiltonian\cite{tony2019} captures the essential physics of the limiting multilayer Lieb ($\theta = \pi /2$) and Kagome ($\theta = 2\pi/3$) systems, as compared throughout the text discussions with the reported results in the literature, demonstrating to be an appropriate and less computationally demanding as compared with first-principles calculations to deal with systems with Lieb, transition, and Kagome crystallographic symmetry under the effect of any external perturbations.

\section*{Acknowledgments}

The authors would like to thank the National Council of Scientific and Technological Development (CNPq) through Universal and PQ programs and the Coordination for the Improvement of Higher Education Personnel (CAPES) of Brazil for their financial support. D.R.C gratefully acknowledges the support from CNPq grants $313211/2021-3$, $437067/2018-1$, $423423/2021-5$, $408144/2022-0$, the Research Foundation—Flanders (FWO - Vl), and the Fundação Cearense de Apoio ao Desenvolvimento Científico e Tecnológico (FUNCAP).

\bibliography{references}
\end{document}